\documentclass[aps, prd, twocolumn, notitlepage, nofootinbib]{revtex4-2}
\usepackage{graphicx}
\usepackage{float}
\usepackage{amsmath}
\usepackage{color}
\usepackage{tensor}
\usepackage{epstopdf}
\usepackage{xcolor}
\usepackage{wasysym}
\usepackage{hyperref}
\hypersetup{
    colorlinks,
    linkcolor={blue!50!black},
    citecolor={red!50!black},
    urlcolor={blue!80!black}
}

\usepackage{multibib}

\newcommand{\be}{\begin{equation}}
\newcommand{\ee}{\end{equation}}
\newcommand{\ba}{\begin{eqnarray}}
\newcommand{\ea}{\end{eqnarray}}

\newcommand{\beq}{\begin{equation}}
\newcommand{\eeq}{\end{equation}}
\newcommand{\beqa}{\begin{eqnarray}}
\newcommand{\eeqa}{\end{eqnarray}}
\newcommand{\nn}{\nonumber}
\graphicspath{{plots/}}

\begin{document}

\title{Lensing Signatures of a  Slowly-Accelerated Black Hole}

\author{Amjad Ashoorioon}
\email{amjad@ipm.ir}
\affiliation{School of Physics, Institute for Research in Fundamental Sciences (IPM), P.O. Box 19395-5531, Tehran, Iran}

\author{Mohammad Bagher Jahani Poshteh}
\email{jahani@ipm.ir}
\affiliation{School of Physics, Institute for Research in Fundamental Sciences (IPM), P.O. Box 19395-5531, Tehran, Iran}

\author{Robert B. Mann}
\email{rbmann@uwaterloo.ca}
\affiliation{Department of Physics and Astronomy, University of Waterloo, Waterloo,
Ontario, N2L 3G1, Canada\\
Perimeter Institute for Theoretical Physics, Waterloo, Ontario, N2L
2Y5, Canada}

%


\begin{abstract}
Accelerating black holes, connected to cosmic strings could evolve to supermassive black holes. However, if they are going to take part in structure formation and resides at the center of galaxies, their acceleration should be small. This slow acceleration does not change the shadow or image position in gravitational lensing effect significantly. However we show that the time delay associated to these images change significantly. This is in contrast with when the theory governing the strong gravitational field around the black hole is different from general relativity, where not only the differential time delays but the angular position of images would be different. We conclude that, if the observed angular position of images are compatible with the prediction of general relativity, a possible deviation in the differential time delays between the observed values and those predicted by general relativity, could be due to the acceleration of the black hole.
\end{abstract}

\maketitle

\section{Introduction}
 
Some gauge theories allow the possibility of topological defects such as cosmic strings in the early Universe. They could be produced at the end of brane inflation~\cite{Sarangi:2002yt} or during first order phase transitions~\cite{Kibble:1976sj,vilenkin1985}. Every  cosmic string has an associated tension, which is given by its mass per  unit length~\cite{vilenkin1985}. Cosmic strings could break or fray to produce black holes ~\cite{hr,eardley1995}. These black holes come in pairs and are accelerating (away from each other) due to the tension of the cosmic strings. Other channels to produce accelerating black holes are through black hole pair production in a background magnetic field~\cite{garfinkle1991,hawking1995,dowker1994,emparan1995}, in de Sitter space~\cite{mellor1989,mann1995,dias2004:2}, or on a cosmic string in background with both positive cosmological constant and magnetic field~\cite{Ashoorioon:2014ipa,ashoorioon2021}.

It is possible that a network of cosmic strings, which gets attached to primordial black holes produced in the early Universe~\cite{vilenkin2018}, exists. Such primordial black holes could form much later than the cosmic string network does; an  example, would be the reentry of fluctuations with large amplitudes generated during inflation \cite{Clesse:2016vqa,Ashoorioon:2019xqc}, or from the bubbles nucleated during inflation and collapse during the ensuing matter or radiation-dominated era \cite{Deng:2016vzb,Ashoorioon:2020hln}.
Such black holes would also be accelerating because of the tension of its attached cosmic string.

Such black holes could serve as seeds for supermassive black holes believed to be at the centers of most galaxies~\cite{vilenkin2018,morris2017}. However, their velocity is limited to be less than ($100\, {\rm km}/{\rm s}$). Otherwise they cannot be captured by host galaxies during   structure formation~\cite{vilenkin2018}. The acceleration of these black holes must therefore be very small: so small that one can assume that a ray of light 
from source to observer that transversally passes the black hole will be  on or near the equatorial plane of the black hole during its motion. In this paper, we study gravitational lensing by such \textit{slowly} accelerating black holes, and we take the acceleration to be perpendicular to the equatorial plane.

In the past few years accelerating black holes have attracted increasing  attention due to their interesting thermodynamic properties~\cite{Astorino:2016ybm,appels2016,anabalon2018,Abbasvandi:2018vsh,Abbasvandi:2019vfz,anabalon2019,Rostami:2019ivr,Ahmed:2019yci}, near horizon symmetries~\cite{brenner2021}, and other features
\cite{Zhang:2019vpf,Astorino:2021rdg}. Null geodesics around these black holes have also been studied ~\cite{lim2021,frost2021} and the shadow cast by them has been investigated in~\cite{grenzebach2015,zhang2021}.

Our aim is to study gravitational lensing by accelerating black holes. Studies of gravitational lensing in the strong field regime of Schwarzchild black holes traces back to the work of Darwin~\cite{Darwin59,Darwin61}. In a lensing effect by a black hole, the latter deflects the nearby light rays  coming from a source, by an angle called the deflection angle. The light rays will then travel toward an observer who cannot see the source in its real place, but sees images of it located elsewhere. The image that is on the same side as the source (with respect to line of sight to the black hole) is called the primary image. There is also an image on the opposite side, which is called the secondary image.

Light rays  passing very close to the black hole circle around the black hole several times before continuing their travel toward the observer. There is one set of an infinite number of such images in either side of the black hole~\cite{Ellis}. Observation of these so-called relativistic images is very hard. Nonetheless if they are observed, they could be used to find very accurate values for the mass of the black hole and its distance to us~\cite{Virbhadra}.

Some features of gravitational lensing by accelerating black holes have been studied for a Carter observer~\cite{frost2021}. Here we study gravitational lensing as seen by a distant static observer,   paying special attention to the time delays associated with the primary and secondary images. We specifically consider the supermassive black hole at the center of the M87 galaxy as the lens and assume its acceleration to be sufficiently small so that a light ray passing it hole stays on or near the equatorial plane during its transit from  source to   observer.  Black holes with such small accelerations can take part in structure formation~\cite{vilenkin2018}. The range of values we assume for  the acceleration parameter is also consistent with   existing bounds  on the abundance of primordial black holes \cite{vilenkin2018,Sasaki:2018dmp}, coming from  dynamical friction on compact objects \cite{Carr:1997cn}.

Recently we proposed a method for determining the acceleration of a black hole by exploiting the fact that differential time delays associated with its lensed images are substantially different with respect to the case of non-accelerating black holes \cite{Ashoorioon:2021znw}. Here expand upon the details of this study and augment the assumption of equatorial motion.

The outline of our paper is as follows. In the next section we present the form of the metric that describes accelerating black holes and we are going to use throughout the paper. In Sec.~\ref{sec:lensing} we review the basic equations of the gravitational lensing. In Sec.~\ref{sec:m87} we investigate gravitational lensing by the black hole at the center of M87 galaxy. We assume this black hole to be non-rotating slowly accelerating as well as non-accelerating and compare the results of primary, secondary and relativistic images. We conclude our paper in Sec.~\ref{sec:con}. We work in geometric units where $G=c=1$ and use the mostly-positive signature for the spacetime metric.

\section{C metric in spherical coordinates}\label{sec:metric}

Spacetime around a black hole with uniform acceleration is described by the  C metric~\cite{kinnersley1970}. An interesting form of this metric is~\cite{hong2003}
\begin{eqnarray}
ds^2&=&\frac{1}{\alpha^2(x+y)^2} \nn\\
&\times&\left[-F(y)d\tau^2+\frac{dy^2}{F(y)}+\frac{dx^2}{G(x)}+G(x)d\phi^2\right],
\label{eqn:metric:hong}
\end{eqnarray}
where
\begin{eqnarray}
F(y)&=&-(1-y^2)(1-2\alpha m y), \nn\\
G(x)&=&(1-x^2)(1+2\alpha m x),
\label{FG}
\end{eqnarray}
and $m$ is the mass parameter. The parameter $\alpha$ is interpreted as the acceleration of the black hole~\cite{griffiths2006}
and can be taken to be positive without loss of generality.
The advantage of this form over the usual representation of the C metric is that   the factorizable form of the structure functions $F(y)$ and $G(x)$ in \eqref{FG} makes the roots trivial to write down. To keep the order of the roots, it is necessary to take $0<2\alpha m<1$. One finds from Eq. \eqref{eqn:metric:hong} that $x+y=0$ corresponds to conformal infinity (see~\cite{hong2003} for more features of this metric).

An issue related to the metric \eqref{eqn:metric:hong} is that it has no convenient limit as $\alpha\rightarrow 0$. To overcome this problemwe employ the transformation~\cite{griffiths2006}
\be
x=\cos \theta, \qquad y=\frac{1}{\alpha r}, \qquad \tau=\alpha t,
\ee
to write C metric in the spherical-like coordinates
\begin{eqnarray}
ds^2&=&\frac{1}{(1+\alpha r\cos\theta)^2} \nn\\
&\times&\left[-Q(r)dt^2+\frac{dr^2}{Q(r)}+\frac{r^2d\theta^2}{P(\theta)}+P(\theta)r^2\sin^2\theta d\phi^2\right], \label{eqn:metric:griffiths}\nn\\
\end{eqnarray}
where
\begin{eqnarray}
Q(r)&=&(1-\alpha^2 r^2)\left(1-\frac{2m}{r}\right), \nn\\
P(\theta)&=&1+2\alpha m \cos\theta. \label{eqn:tran:griffiths}
\end{eqnarray}
It is obvious that the above metric reduces to the Schwarzschild solution\footnote{The general case of charged rotating black holes, which also possess a NUT charge parameter, was obtained in~\cite{griffiths2005}.} for $\alpha\rightarrow 0$.

The Kretschmann curvature invariant obtained from the Riemann tensor $R_{\mu\nu\gamma\delta}$, for the metric \eqref{eqn:metric:griffiths}, is
\be
R_{\mu\nu\gamma\delta}R^{\mu\nu\gamma\delta}=48 m^2 \left(\frac{1}{r}+\alpha  \cos \theta\right)^6=48 m^2 \alpha^6(x+y)^6, \label{eqn:cur:griffiths}
\ee
which diverges at $r=0$, indicating a physical singularity at this point. Note that~\eqref{eqn:cur:griffiths} yields the Kretschmann scalar for a Schwarzschild black hole in the $\alpha\rightarrow 0$ limit, and that for $m=0$ it vanishes, indicating a flat spacetime.

Eq. \eqref{eqn:cur:griffiths} shows that the spacetime is also flat as $x+y\rightarrow 0$, or in other words near conformal infinity. However, the curvature invariant does not vanish for $r\rightarrow\infty$ unless $\theta=\pi/2$, which is a disadvantage of the form \eqref{eqn:metric:griffiths} of the C metric. Fortunately we need not  be concerned about this problem since we would like to study the lensing effects on the equatorial plane of accelerating black holes, $\theta=\pi/2$. On this plane, the metric \eqref{eqn:metric:griffiths} can be written as
\be
ds^2=-Qdt^2+\frac{dr^2}{Q}+r^2d\phi^2, \label{eqn:metric}
\ee
with $Q(r)$ given by Eq.~\eqref{eqn:tran:griffiths}.

\section{Gravitational lensing by black holes}\label{sec:lensing}

\begin{figure}[htp]
	\centering
	\includegraphics[width=0.5\textwidth]{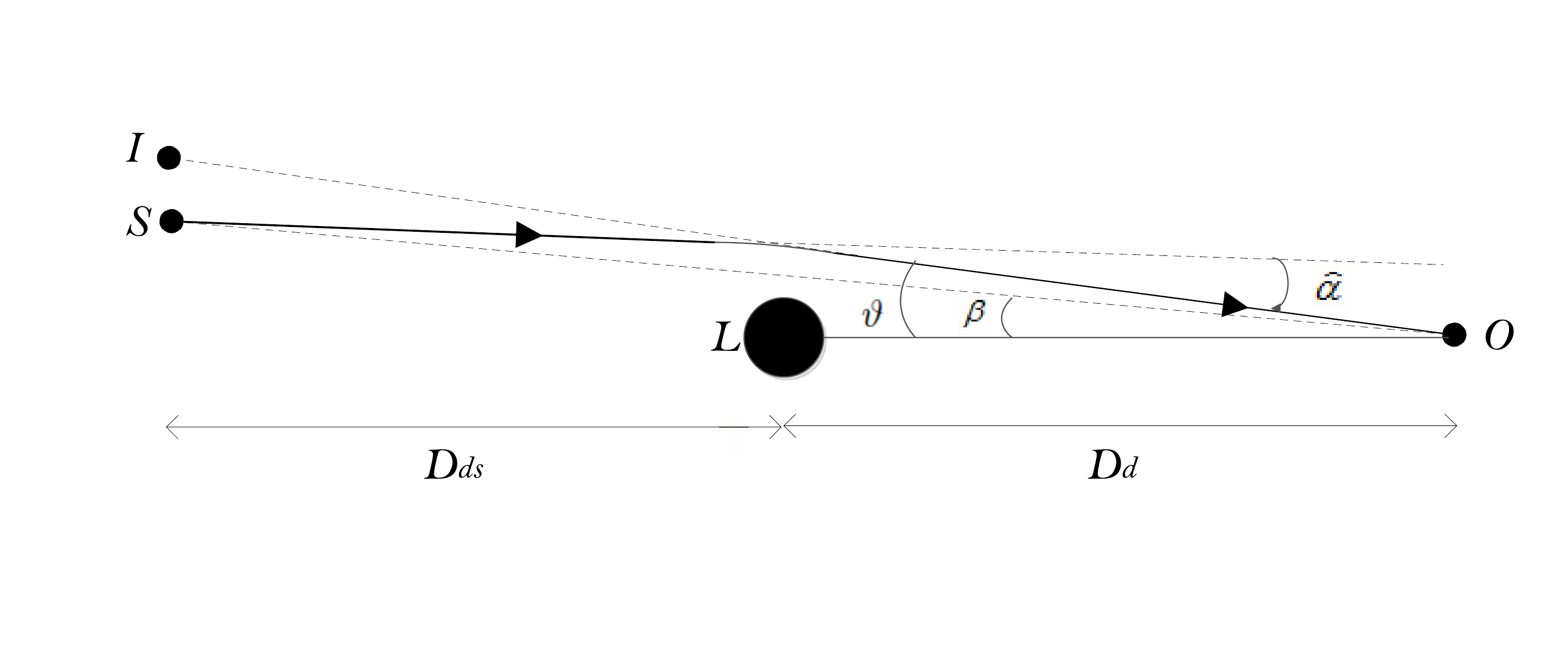}
	\caption{{\it Deflection of light ray by the black hole}: The source, image, observer, and the black hole (lens) are shown by $S$, $I$, $O$, and $L$, respectively. $D_d$ is the distance from the black hole to the observer and $D_{ds}$ is the distance from the black hole to the source. We take the plane of this figure to be the equatorial plane of the black hole. We assume that the acceleration $\alpha$ of the black hole is in a direction perpendicular to the plane of this figure and behind it there is an acceleration horizon at $r=1/\alpha$. The sign of $\alpha$ is not important in our approximation. $\beta$ shows the actual angular position of the source with respect to the line of sight to the black hole. The black hole deflect the ray of light by deflection angle $\hat{\alpha}$ so that the observer sees the image of the source with the angular position $\vartheta$.}
	\label{fig:lens_diag}
\end{figure}

In this section we review the basic equations that describe the gravitational lensing by black holes. Fig.~\ref{fig:lens_diag} schematically shows the effect that we are going to study in this paper. The black hole deflects  light rays that are coming from a source $S$ by a deflection angle $\hat{\alpha}$. The observer $O$ then sees the image $I$ of the object with angular position $\vartheta$.

Geodesic motion on the equatorial plane of the black hole is described by the Lagrangian
\be
\mathcal{L}=\frac{1}{2}g_{\mu\nu}\dot{x}^\mu\dot{x}^\nu=\frac{1}{2}\left(-Q\dot{t}^2+\frac{\dot{r}^2}{Q}+r^2\dot{\phi}^2\right), \label{eqn:lag}
\ee
with dot denoting differentiation with respect to some affine parameter along the geodesic. Since  the particle is assumed to be on the equatorial plane
we set $\dot{\theta}=0$; we shall discuss this approximation below.
 The constants of motion are
\be
E=-\frac{\partial \mathcal{L}}{\partial \dot{t}}=Q\dot{t},   \qquad L_z=-\frac{\partial \mathcal{L}}{\partial \dot{\phi}}=-r^2\dot{\phi}.
\ee

For the null geodesics  $\mathcal{L}=0$ and we  find from Eq.~\eqref{eqn:lag}
\be\label{eqn:lagf1}
\frac{1}{Qr^2}\left(\frac{dr}{d\phi}\right)^2=\frac{r^2}{Q}\frac{E^2}{L_z^2}-1\; .
\ee
At the point of closest approach to the black hole, $r=b$, we have $\frac{dr}{d\phi}=0$. Eq. \eqref{eqn:lagf1}   then yields $E^2/L_z^2=Q_b/b^2$, where $Q_b=Q(r=b)$. Consequently Eq.~\eqref{eqn:lagf1} can be written as
\be\label{eqn:dphidr}
\frac{d\phi}{dr}=\frac{1}{r\sqrt{\left(\frac{r}{b}\right)^2Q_b-Q}}.
\ee
The deflection angle can be found by\footnote{It is obvious from Eq.~\eqref{eqn:tran:griffiths} that the metric function $Q$ changes sign at $r=1/\alpha$. Due to the factor $1/r$ in Eq.~\eqref{eqn:alpha_hat}, the large $r$ contribution from $\alpha^{-1}$ to infinity is negligible in the integral of \eqref{eqn:alpha_hat} (about one part in a million). Therefore it is safe to integrate to infinity.}~\cite{weinberg1972}
\be\label{eqn:alpha_hat}
\hat{\alpha}(b)=2\int_{b}^{\infty}\frac{dr}{r\sqrt{\left(\frac{r}{b}\right)^2Q_b-Q}}-\pi,
\ee
where we have assumed that the distance from the black hole to the observer, $D_d$, and the distance from the black hole to the source, $D_{ds}$, are both much larger than $b$. We note that although the distances from the black hole to the observer/source are  very large compared to the radius of closest approach, they are still much smaller than the length scale set by the acceleration, i.e.~$1/\alpha\gg D_d, D_{ds}\gg b$. In this regime, on the equatorial plane, the line element \eqref{eqn:metric} is the appropriate one to work with.

Note that we have taken the direction of the acceleration to be perpendicular to the equatorial plane of the black hole (plane of Fig.~\ref{fig:lens_diag}). For situations in which the acceleration has a component parallel to this plane, the third component of the angular momentum is not constant and the geodesic equations cannot be integrated analytically.

The time delay is the difference between the time it takes for the light to travel the physical path from the source to the observer and the time it takes to travel the path from the source to the observer when there is no black hole. To find the time delay we first rewrite the Lagrangian \eqref{eqn:lag} for  null geodesics as
\be
\frac{1}{Q^2}\left(\frac{dr}{dt}\right)^2=1-\dfrac{Q}{r^2}\frac{L_z^2}{E^2}.
\ee
Since $\frac{dr}{dt}=0$ at $r=b$, we obtain
\be\label{eqn:dtdr}
\frac{dt}{dr}=\frac{1}{Q\sqrt{1-\left(\frac{b}{r}\right)^2\frac{Q}{Q_b}}}.
\ee
The time delay can then be found by the integral
\be\label{eqn:time_delay}
\tau(b)=\left[\int_{b}^{r_s}dr+\int_{b}^{D_d}dr\right]\frac{1}{Q\sqrt{1-\left(\frac{b}{r}\right)^2\frac{Q}{Q_b}}}-D_s\sec\beta,
\ee
in which $\beta$ is the angular position of the source, $D_s=D_d+D_{ds}$ is the distance from the source to the observer, and $r_s=\sqrt{D_{ds}^2+D_s^2\tan^2\beta}$.

The image angular position, $\vartheta$, obeys the Virbhadra-Ellis lens equation~\cite{Ellis}
\be\label{eqn:lens_eq}
\tan\beta=\tan\vartheta-\mathcal{D}\left[\tan\vartheta+\tan(\hat{\alpha}-\vartheta)\right],
\ee
where $\mathcal{D}=D_{ds}/D_s$. The impact parameter is given by~\cite{virbhadra1998}
\be\label{eqn:impact}
J=\frac{b}{\sqrt{Q_b}}=D_d\sin\vartheta.
\ee
Also the image magnification is
\be\label{eqn:maggi}
\mu=\left(\frac{\sin\beta}{\sin\vartheta}\frac{d\beta}{d\vartheta}\right)^{-1}.
\ee
Differentiation of Eq. \eqref{eqn:lens_eq} with respect to $\vartheta$ yields
\be
\sec^2\beta\frac{d\beta}{d\vartheta}=\sec^2\vartheta-\mathcal{D}\left[\sec^2\vartheta+\sec^2(\hat{\alpha}-\vartheta)\left(\frac{d\hat{\alpha}}{d\vartheta}-1\right)\right].
\ee
To find the derivative of the deflection angle with respect to $\vartheta$, we use $\frac{d\hat{\alpha}}{d\vartheta}=\frac{d\hat{\alpha}}{db}\frac{db}{d\vartheta}$. The factor $\frac{db}{d\vartheta}$ can be found from Eq. \eqref{eqn:impact}. The factor $\frac{d\hat{\alpha}}{db}$ is a bit tricky and has been found in~\cite{poshteh2019} as
\be\label{eqn:alpha_r}
\frac{d\hat{\alpha}(b)}{db}=-2\int_{b}^{\infty}\frac{1}{\sqrt{\mathcal{F}}}\frac{\partial}{\partial r}\left(\frac{1}{r}\frac{\partial\mathcal{F}}{\partial b}\frac{\partial r}{\partial\mathcal{F}}\right)dr,
\ee
where $\mathcal{F}=\left(\frac{r}{b}\right)^2Q_b-Q$. These results will be used in next section to study gravitational lensing by M87* in both accelerating and non-accelerating cases.

We have assumed that the light ray is in the plane  $\theta = \pi/2$, but this surface is not a geodesic surface.   To check the validity of
this approximation, we write  $\theta = \pi/2 + \delta$. We can assume without loss of generality that the light ray begins with $\theta = \pi/2$. 
The direction of the acceleration is perpendicular to the equatorial plane of the black hole and the geodesic equations indicate that the exterior curvature of the $\theta = \pi/2$ surface is non-vanishing \cite{griffiths2006}. However for small acceleration, the 
deviation of the
path of the light ray   from the equatorial plane is $\delta\sim \frac{\alpha D_s^2}{D_s}=\alpha D_s$, regardless of the cause of the acceleration. 
Since both source and  observer are assumed to be inside the acceleration horizon, $\alpha D_s \ll 1$. 
This tiny deviation changes the time delays by a small fraction $\sim \alpha^2 D_s^2$, and so our results  for   differential time delays will be valid up to order $\alpha^2 D_s^2$.  Furthermore, $\dot{\delta} = \dot{\theta} \sim \theta/t \sim \alpha D_s/D_s \sim \alpha\ll D_s^{-1}$, and so   $\delta\equiv \theta-\frac{\pi}{2}\ll 1$ throughout the motion.


To see how a small value of $\delta$ changes the deflection angle and time delay, for small $\delta$ we have
\begin{eqnarray}
	\sin \left(\frac{\pi}{2}+\delta\right) &\simeq& 1 +\mathcal{O}(\delta^2),\\
	\cos \left(\frac{\pi}{2}+\delta\right) &\simeq& -\delta +\mathcal{O}(\delta^3).
\end{eqnarray}
and within this approximation  the metric is
\begin{eqnarray}
	ds^2&=&\frac{1}{(1-\alpha r \delta)^2} \nn\\
	&\times&\left[-Q(r)dt^2+\frac{dr^2}{Q(r)}+\frac{r^2 d\delta^2}{P(\delta)}+P(\delta)r^2 d\phi^2\right], \label{eqn:metric:griffiths:del}\nn\\
\end{eqnarray}
where
\begin{eqnarray}
	Q(r)&=&(1-\alpha^2 r^2)\left(1-\frac{2m}{r}\right), \nn\\
	P(\delta)&=&1-2\alpha m \delta. \label{eqn:tran:griffiths:del}
\end{eqnarray}

Dropping the factor $\dot{\delta}^2$, which as argued is quite small, the equations governing the geodesics can be obtained using the Lagrangian
\be
\mathcal{L}=\frac{1}{2}g_{\mu\nu}\dot{x}^\mu\dot{x}^\nu=\frac{1}{2\left(1-\alpha r \delta\right)^2}\left(-Q\dot{t}^2+\frac{\dot{r}^2}{Q}+Pr^2\dot{\phi}^2\right), \label{eqn:lag:del}
\ee
The constants of motion are
\begin{eqnarray}
	E&=&-\frac{\partial \mathcal{L}}{\partial \dot{t}}=\frac{Q\dot{t}}{(1-\alpha r \delta)^2},\nn\\
	L_z&=&-\frac{\partial \mathcal{L}}{\partial \dot{\phi}}=\frac{-Pr^2\dot{\phi}}{(1-\alpha r \delta)^2}.
\end{eqnarray}
Now we can write the Lagrangian for null geodesics as
\be
\mathcal{L}=\frac{1}{2\left(1-\alpha r \delta\right)^2}\left[\frac{1}{Q P r^2}\left(\frac{dr}{d\phi}\right)^2-\frac{E^2r^2 P}{Q L_z^2}+1\right]=0,
\ee
from which we find
\be
\frac{1}{Q P r^2}\left(\frac{dr}{d\phi}\right)^2=\frac{E^2r^2 P}{Q L_z^2}-1.
\ee

At closest approach $r = b$, we have $dr/d\phi = 0$, and one finds
\be
\frac{E^2}{L_z^2} = \frac{Q_b}{b^2 P},
\ee
where $Q_b = Q(r=b)$. Using the above equation we find
\be
\frac{d\phi}{dr} = \frac{\left(1-2\alpha m \delta\right)^{-1/2}}{r\sqrt{\left(\frac{r}{b}\right)^2Q_b-Q}}
\ee
and so the deflection angle is
\be\label{eqn:alpha_hat:del}
\hat{\alpha}(b) \simeq 2\int_{b}^{\infty}\frac{1+\alpha m \delta}{r\sqrt{\left(\frac{r}{b}\right)^2Q_b-Q}}dr-\pi.
\ee
The first order correction to $\hat{\alpha}$ is very small; in the example considered in this paper we have $\alpha m = \mathcal{O}(10^{-12})$. There is further suppression due to the factor $\delta$ which, as we have shown above, is $\mathcal{O}(\alpha D_s)\sim 10^{-3}$.

The Lagrangian \eqref{eqn:lag} can also be written as
\be
\mathcal{L}=\frac{1}{2\left(1-\alpha r \delta\right)^2}\left(\frac{\dot{r}^2}{Q^2\dot{t}^2}+\frac{Pr^2\dot{\phi}^2}{Q\dot{t}^2}-1\right).
\ee
For null geodesics $\mathcal{L} = 0$, therefore
\be
\frac{1}{Q^2}\left(\frac{dr}{dt}\right)^2=1-\frac{Q}{r^2P}\frac{L_z^2}{E^2}.
\ee
At $r=b$ we have $L_z^2/E^2 = b^2 P/Q_b$, so
\be
\frac{1}{Q^2}\left(\frac{dr}{dt}\right)^2=1-\frac{Q}{r^2}\frac{b^2}{Q_b}.
\ee
$\delta$ does not appear in the above equation, which means that Eq. \eqref{eqn:time_delay} given for the time delay does not change to first order in $\delta$. Indeed the leading order correction to the time delay is of order $\delta^2$.

\section{Gravitational lensing by M87*}\label{sec:m87}

In this section we study the gravitational lensing by the black hole at the center of the M87 galaxy. Since zero rotation is allowed by Event Horizon Telescope observation of M87*, we take this black hole to be non-rotating~\cite{EventHorizonTelescope:2021dqv}.
We shall employ the numerical methods of~\cite{Ellis,poshteh2019} to study gravitational lensing by M87*,  assuming this black hole is accelerating and compare the results to the case in which it is non-accelerating. Our study includes investigation of image positions, magnifications, and (differential) time delays associated to images in a gravitational lensing effect.

The mass of M87* and our distance to it have been obtained by Event Horizon Telescope Collaboration as $M_{{\rm M87^*}}=9.6\times 10^{12} \, {\rm m}\equiv 6.5\times 10^9 M_{\odot}$ , $D_d=5.2\times 10^{23} \, {\rm m}$~\cite{Akiyama:2019eap}. We take the acceleration 
of the black hole to be $\alpha=10^{-25} {\rm m}^{-1}$, roughly the upper bound obtained by \cite{vilenkin2018} from the development of a cosmic string network where we take black holes of such  masses as beads in the network. This is small 
 enough to ensure that a light ray stays roughly on the equatorial plane throughout its passage from  source to observer.   

We also assume that the velocity of the black hole has remained small enough to contribute to structure formation and the relative abundance of such black holes to the dark matter energy density is consistent with   dynamical constraints coming from the disruption of binary stars or open star clusters, $f_{{}_{\rm PBH}}\lesssim 10^{-3}$, \cite{Carr:1997cn}. We also assume that the black hole is halfway between the source and the observer; therefore ${\cal D}=0.5$.

\begin{figure}[htp]
	\centering
	\includegraphics[width=0.5\textwidth]{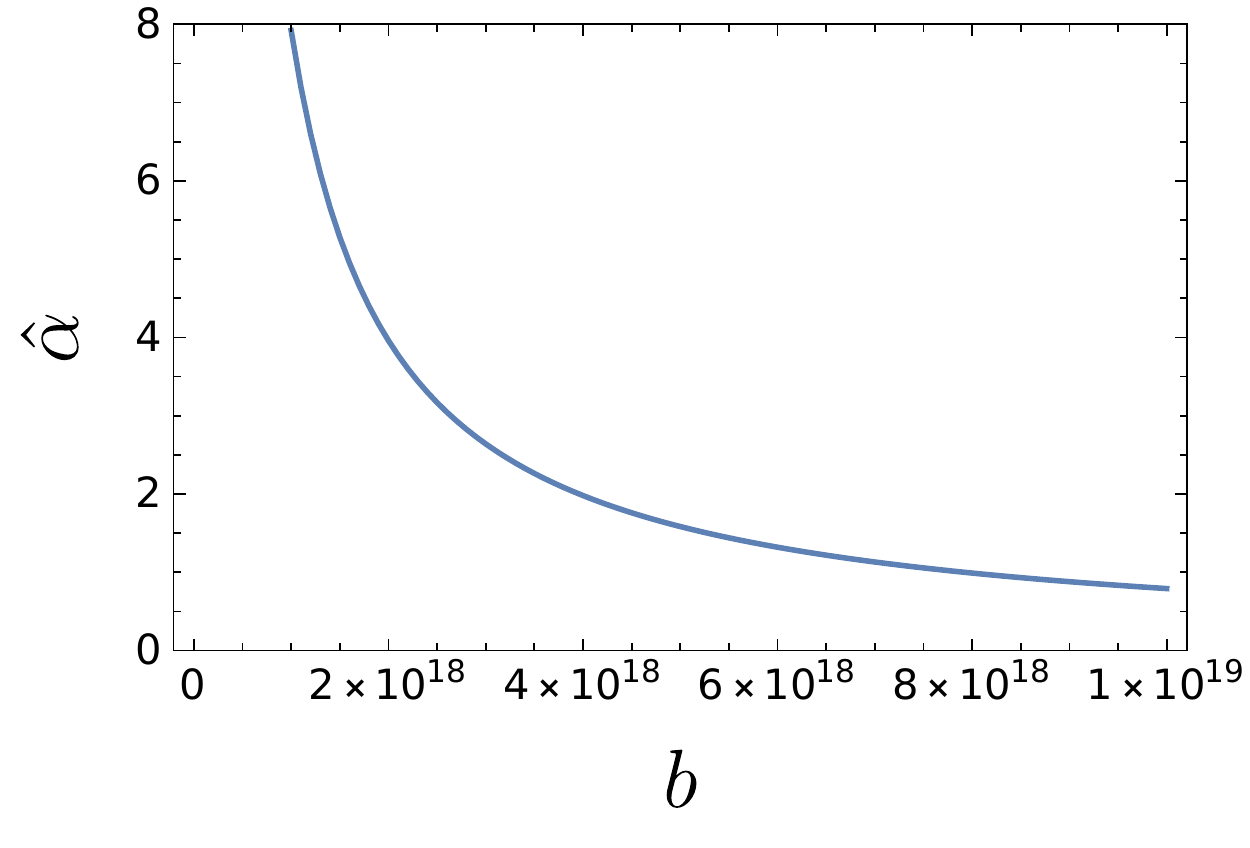}
	\caption{{\it Deflection angle as a function of impact parameter}: The deflection angle $\hat{\alpha}$ is in {\em arcseconds} and the impact parameter $b$ is in {\em meters}. We have set $M_{{\rm M87^*}}=9.6\times 10^{12} \, {\rm m}$ and $\alpha=10^{-25} {\rm m}^{-1}$.}
	\label{fig:defang_vs_b}
\end{figure}

In Fig.~\ref{fig:defang_vs_b} we  use~\eqref{eqn:alpha_hat} to plot the deflection angle $\hat{\alpha}$ as a function of the impact parameter $b$,
taking the black hole mass to be that of  M87* of slow acceleration $\alpha=10^{-25} {\rm m}^{-1}$, and  assuming the distance from the black hole to the observer/source is much larger that the impact parameter.   We see that the closer the light ray is as it passes the black hole, the larger the  deflection angle.  For the same value of the impact parameter, the deflection angle of the $\alpha=0$ non-accelerating black hole is larger than the deflection angle of accelerating black hole only by about 1 part in $10^{13}$. Such deviations cannot be observed in the near future.

\subsection{Primary and secondary images}

In Table~\ref{tab:psimg}, we have used Eqs.~\eqref{eqn:alpha_hat}, \eqref{eqn:lens_eq}, \eqref{eqn:impact}, \eqref{eqn:maggi}, and \eqref{eqn:alpha_r}, to compute image positions $\theta$, deflection angles $\hat{\alpha}$, impact parameters $b$, and magnifications $\mu$ of primary and secondary images for different values of angular positions of the source $\beta$. The values of these parameters are nearly the same for a non-accelerating and slowly accelerating M87*, i.e.~they do not depend on the acceleration significantly. For a fixed value of $\beta$, the impact parameter in the non-accelerating case is greater than that of the accelerating case by only about 1 part in $10^{13}$ parts. Meanwhile the (absolute value of) the image angular position is larger for the accelerating case by only about 1 part in $10^{17}$.

Consequently, from current observational technology it is impossible to tell if M87* is non-accelerating or slowly accelerating by using the image positions produced in gravitational lensing.  Likewise,
as we will show in Appendix \ref{app:shadow},  it is also impossible to tell whether or not M87* is  slowly accelerating by using its shadow. The (absolute values of)   magnification of images are also larger in the slowly accelerating case by about 1 part in $10^{17}$.
One might therefore expect that determination of the acceleration of M87* is not possible; as we shall demonstrate below, this is not the case.

\begin{figure}[htp]
	\centering
	\includegraphics[width=0.5\textwidth]{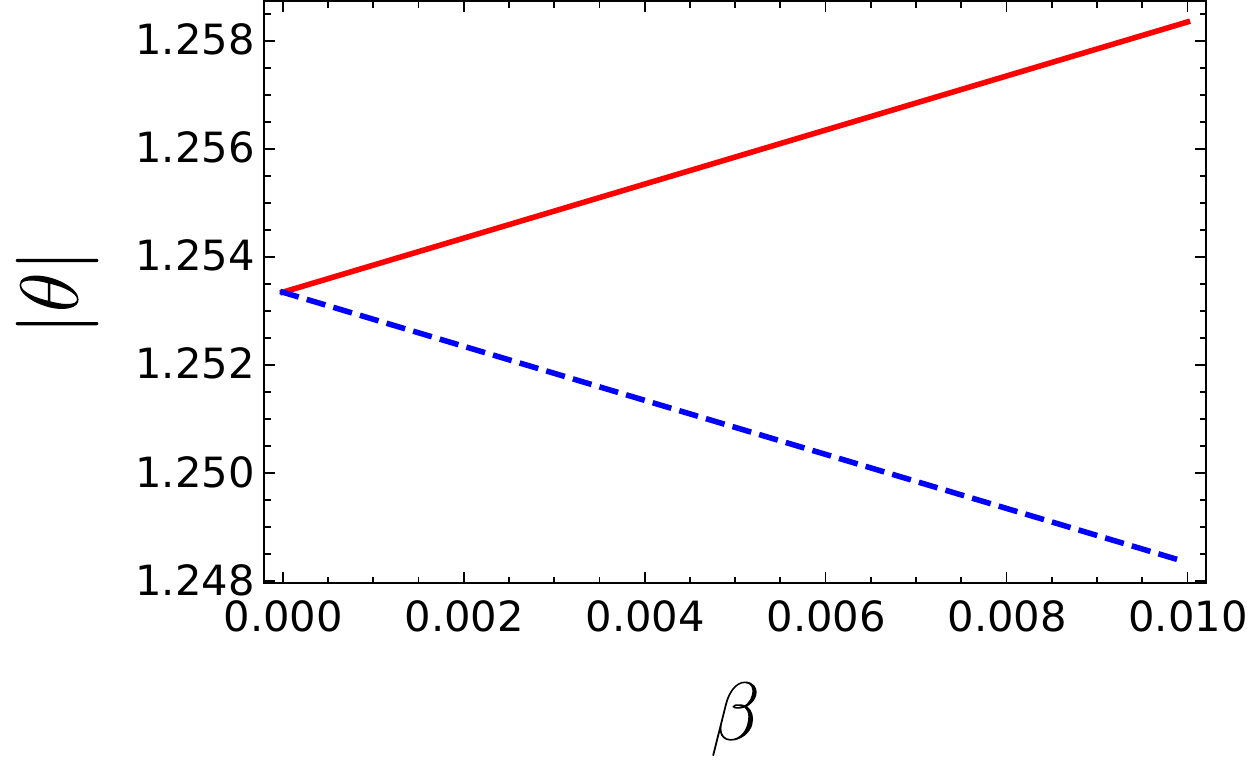}
	\includegraphics[width=0.5\textwidth]{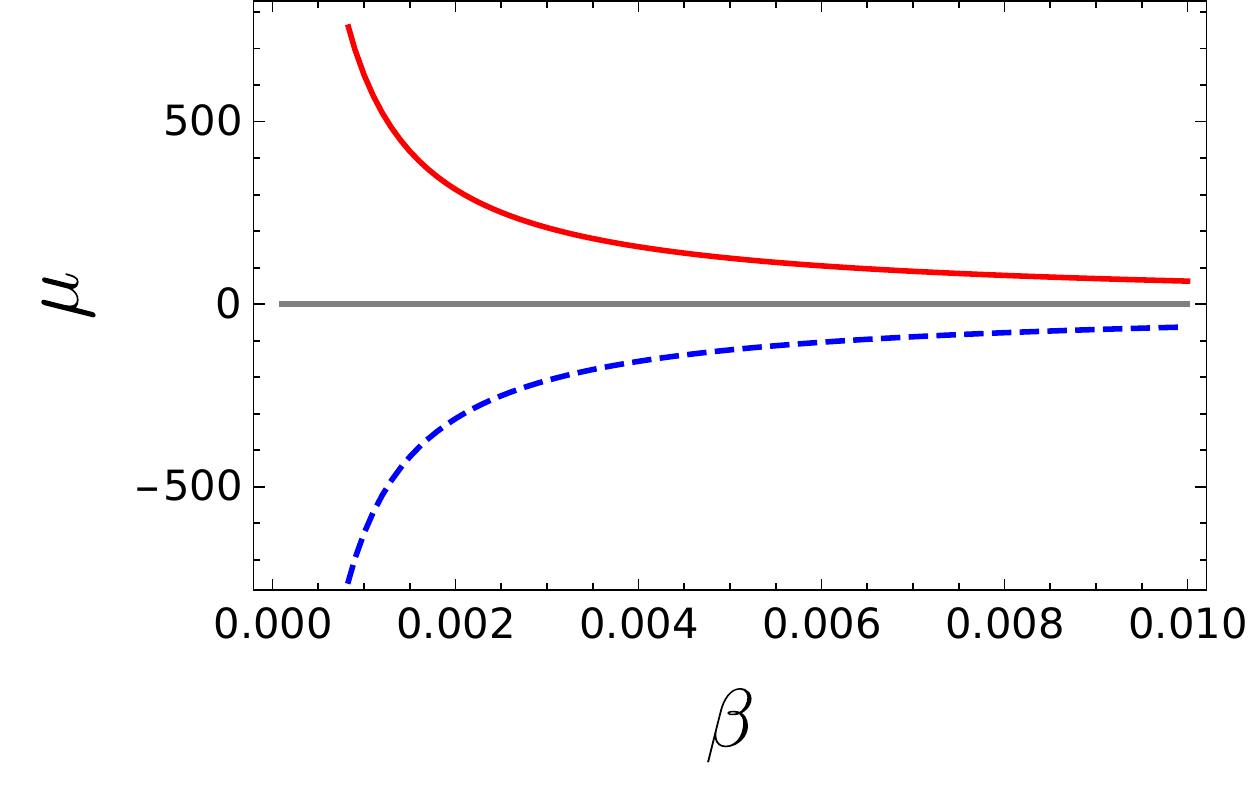}
	\caption{{\it Image position and magnification of primary and secondary images}: (Top) Angular position of the primary images $\theta_{p}$ (red line) and the absolute value of angular position of the secondary images $|\theta_{s}|$ (dashed blue line) as a function of angular position of the source $\beta$. (Bottom) Magnification of primary (red line) and secondary (dashed blue line) images. Angles are in {\em arcseconds}. We have taken $M_{{\rm M87^*}}=9.6\times 10^{12} \, {\rm m}$, $D_d=5.2\times 10^{23} \, {\rm m}$, ${\cal D}=0.5$, and $\alpha=10^{-25} {\rm m}^{-1}$.}
	\label{fig:theta_mu_vs_beta}
\end{figure}

In Table~\ref{tab:psimg} we list the mage positions, deflection angles, impact parameters, and magnifications of primary and secondary images due to lensing by M87*. We find that the angular position of the primary images increases with increasing angular position of the source. However, the absolute value of the angular positions of secondary images {\it decreases} with increasing   angular source position. Furthermore, the absolute values of the magnifications of primary and secondary images decrease with increasing   angular source position. This means that it would be more difficult to observe sources with larger $\beta$. We illustrate these results in Fig.~\ref{fig:theta_mu_vs_beta}.

\begingroup
\begin{table*}
	\caption{{\it Image positions, deflection angles, impact parameters, and magnifications of primary and secondary images due to lensing by M87*}: Angular positions $\theta$, bending angles $\hat{\alpha}$, impact parameters $b$, and magnifications $\mu$ are given for different values of source angular position $\beta$. These results are the same in the non-accelerating and slowly accelerating cases. (a) $p$ and $s$ refer to primary and secondary images, respectively. (b) Angles are in {\em arcseconds} and the impact parameters are in {\em meters}. (c) We have used $M_{{\rm M87^*}}=9.6\times 10^{12} \, {\rm m}$, $D_d=5.2\times 10^{23} \, {\rm m}$, ${\cal D}=0.5$, and $\alpha=10^{-25} {\rm m}^{-1}$.}\label{tab:psimg}
	\begin{ruledtabular}
		\begin{tabular}{l cccc cccc}
			 $\beta$&$\theta_{p}$&$\hat{\alpha}_{p}$&$b_p$&$\mu_{p}$&$\theta_{s}$&$\hat{\alpha}_{s}$&$b_s$&$\mu_{s}$\\
			\hline
			$0$&$1.25334$&$2.50668$&$3.20\times 10^{18}$&$\times$&$-1.25334$&$2.50668$&$3.20\times 10^{18}$&$\times$\\
			$0.1$&$1.30436$&$2.40872$&$3.29\times 10^{18}$&$6.79185$&$-1.20436$&$2.60872$&$3.04\times 10^{18}$&$-5.78173$\\
			$0.5$&$1.52806$&$2.05611$&$3.85\times 10^{18}$&$1.82812$&$-1.02805$&$3.05611$&$2.59\times 10^{18}$&$-0.826954$\\
			$1$&$1.84943$&$1.69886$&$4.66\times 10^{18}$&$1.26694$&$-0.849483$&$3.69897$&$2.14\times 10^{18}$&$-0.267388$\\
			$2$&$2.60342$&$1.20683$&$6.56\times 10^{18}$&$1.05575$&$-0.603448$&$5.20690$&$1.51\times 10^{18}$&$-0.0567785$\\
			$3$&$3.45471$&$0.909423$&$8.71\times 10^{18}$&$1.01705$&$-0.454745$&$6.90949$&$1.14\times 10^{18}$&$-0.0176322$\\
			$4$&$4.36033$&$0.720660$&$1.10\times 10^{19}$&$1.00676$&$-0.360286$&$8.72057$&$9.08\times 10^{17}$&$-0.00687431$\\
		\end{tabular}
	\end{ruledtabular}
\end{table*}
\endgroup

\begingroup
\begin{table*}
	\caption{{\it Time delays of primary and secondary images due to lensing by M87*}: Time delays $\tau$ are given for different values of source angular position $\beta$. Instead of the time delays of secondary images $\tau_s$, we present the differential time delay $t_d=\tau_s-\tau_p$ which is of observational importance. (a) As in Table~\ref{tab:psimg}. (b) $\beta$ is in {\em arcseconds} and the (differential) time delays are in {\em seconds}. (c) As in Table~\ref{tab:psimg}. (d) Barred quantities refer to values of the case that the black hole is not accelerating. We define $\Delta t_d=\bar{t}_d-t_d$.}\label{tab:pst}
	\begin{ruledtabular}
		\begin{tabular}{l cc ccc}
			$\beta$&$\tau_{p}$&$t_d$&$\bar{\tau}_p$&$\bar{t}_d$&$\Delta t_d$\\
			\hline
			$0$&$3.13186970\times 10^{12}$&$0$&$1.62727362\times 10^{6}$&$0$&$0$\\
			$0.1$&$3.13186969\times 10^{12}$&$20445.4686829920$&$1.61725037\times 10^{6}$&$20445.4686829935$&$1.52795\times 10^{-9}$\\
			$0.5$&$3.13186965\times 10^{12}$&$102871.116191642$&$1.58092945\times 10^{6}$&$102871.116191651$&$8.42556\times 10^{-9}$\\
			$1$&$3.13186961\times 10^{12}$&$209681.897709557$&$1.54280388\times 10^{6}$&$209681.897709578$&$2.17115\times 10^{-8}$\\
			$2$&$3.13186955\times 10^{12}$&$448736.764965694$&$1.48443381\times 10^{6}$&$448736.764965783$&$8.89995\times 10^{-8}$\\
			$3$&$3.13186951\times 10^{12}$&$737888.423902370$&$1.44178393\times 10^{6}$&$737888.423902647$&$2.76603\times 10^{-7}$\\
			$4$&$3.13186948\times 10^{12}$&$1089214.16844197$&$1.40880628\times 10^{6}$&$1089214.168442669$&$7.01752\times 10^{-7}$\\
		\end{tabular}
	\end{ruledtabular}
\end{table*}
\endgroup

In Table~\ref{tab:pst} we list the time delays of primary images due to lensing by M87*  using~\eqref{eqn:time_delay} for both non-accelerating
(barred) and accelerating (unbarred) cases, with acceleration $\alpha=10^{-25} {\rm m}^{-1}$ in the latter case. In both cases the time delay of a primary image decreases with increasing   angular source position, whereas for the secondary images,  we can show that the time delay increases with increasing $\beta$. It is very interesting that even this small value of the acceleration increases the time delay by 6 orders of magnitude, notwithstanding that the deflection angles do not differ significantly if the black hole is accelerating. 

The reason for this is as follows. It is obvious from~\eqref{eqn:tran:griffiths} that the acceleration makes significant changes to the metric function only at large distances. However large distances do not provide a significant contribution in the integral of Eq.~\eqref{eqn:alpha_hat} due to the presence of the $1/r$ factor. Therefore, the deflection angle is nearly insensitive to the value of the acceleration. However for the time delay the $1/r$ factor is absent in the integrals of~\eqref{eqn:time_delay} and so large distances make a significant contribution.  Consequently small acceleration can significantly change the time delay.

\begin{figure}[htp]
	\centering
	\includegraphics[width=0.5\textwidth]{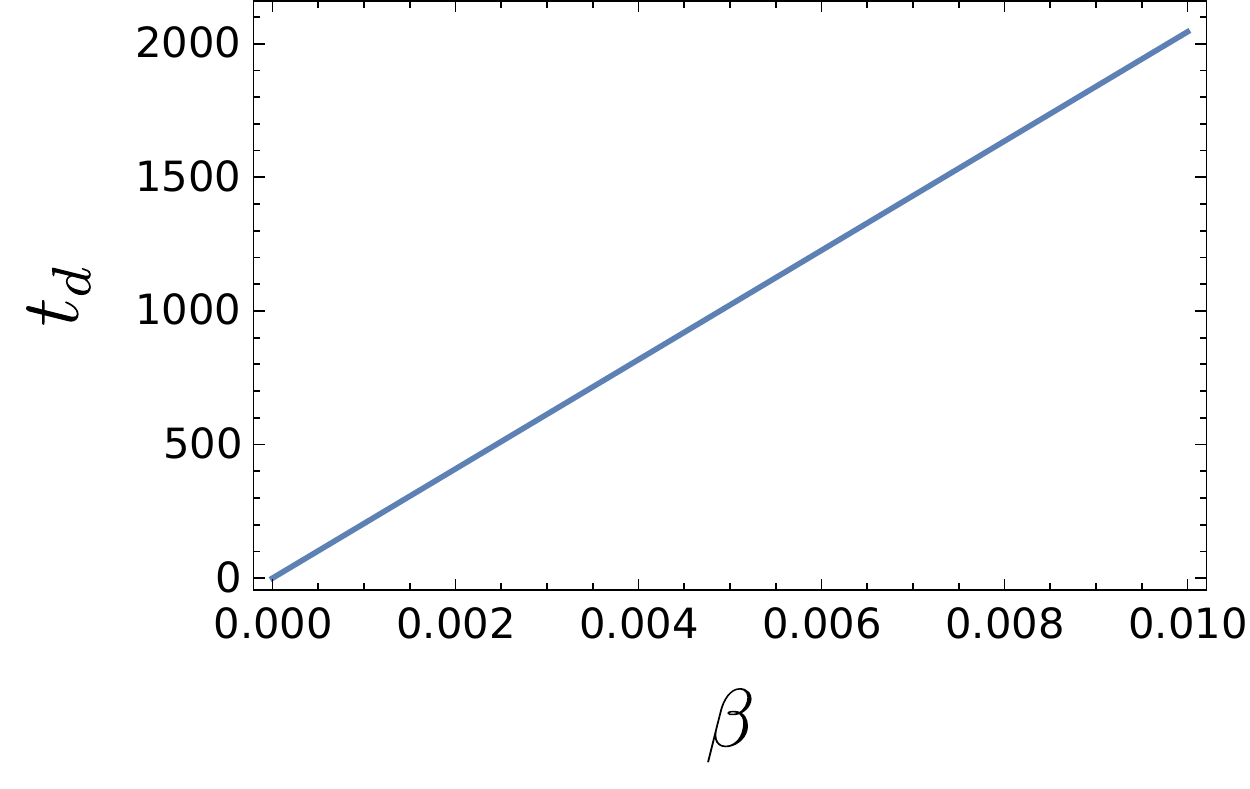}
	\includegraphics[width=0.5\textwidth]{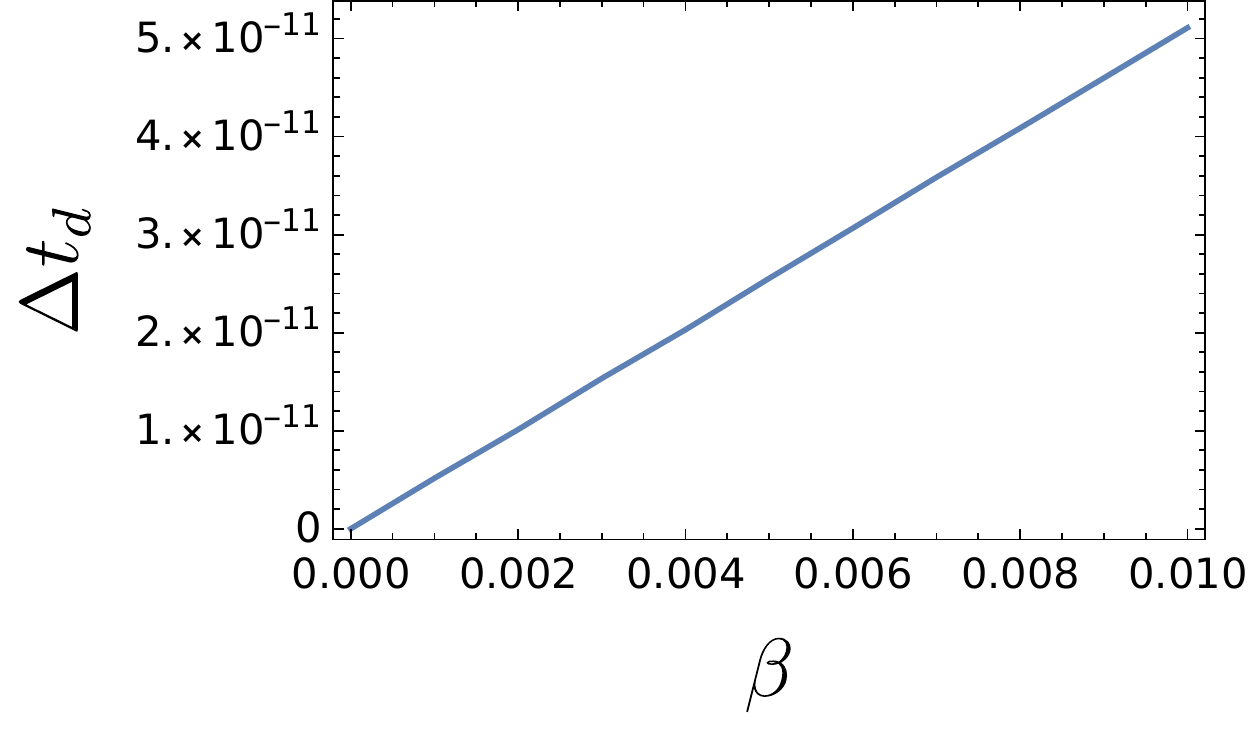}
	\caption{{\it Differential time delay}: (Top) The differential time delay $t_d=\tau_s-\tau_p$ of secondary and primary images as a function of source angular position for accelerating black hole. (Bottom) The difference between the differential time delay in non-accelerating and accelerating cases, $\Delta t_d=\bar{t}_d-t_d$. Time delays are in units of {\em seconds} and $\beta$ is in units of {\em arcseconds}. We have taken $M_{{\rm M87^*}}=9.6\times 10^{12} \, {\rm m}$, $D_d=5.2\times 10^{23} \, {\rm m}$, ${\cal D}=0.5$, and $\alpha=10^{-25} {\rm m}^{-1}$.}
	\label{fig:t_d_vs_beta}
\end{figure}

The time delay, itself, is not an observable. What is observable is the differential time delay $t_d=\tau_s-\tau_p$ (and $\overline{t}_d$). For a pulsating source, every phase in its period  appears in the secondary image $t_d$ seconds after it appears in the primary image. We provide the differential time delay in Table~\ref{tab:pst} (we do not present the explicit values of time delays of secondary images). We find that the differential time delay $t_d$ increases with increasing angular position of  the source, as shown in the top panel of Fig.~\ref{fig:t_d_vs_beta}.

It is interesting that, although the acceleration changes the values of $\tau_p$ (and $\tau_s$) significantly (6 orders of magnitude in our example), the observable quantity $t_d$ does not deviate from its non-accelerating counterpart $\bar{t}_d$ that much. Still the difference is much larger than
what one would naively expect from expanding the integrand in equation \eqref{eqn:time_delay}
to order $\alpha^2$.
\footnote{Expanding the integrand in Eq. \eqref{eqn:time_delay} for the time delay, the coefficient of first correction to the Schwarzschild value, which is proportional to  $\alpha^2$,   turns out to be large and compensates for the smallness of $\alpha^2$, even for small values of the acceleration parameters. Nonetheless, the value of the first order correction is not the same as the integral of the exact integrand as a function of $\alpha$, since one has to sum up all orders of $\alpha$.}
In Table~\ref{tab:pst} we also have presented the difference between the differential time delay in non-accelerating case and the differential time delay in accelerating case $\Delta t_d=\bar{t}_d-t_d$. For primary and secondary images this quantity is positive, which means that for a fixed value of angular source position, the differential time delay is larger if the black hole is not accelerating. The difference increases by increasing the angular source position as   also shown in the bottom panel of Fig.~\ref{fig:t_d_vs_beta}.

These results show that it is indeed feasible to observe if M87* is accelerating or not, provided small changes $\Delta t_d$ in the value of the differential time delay can be measured. We note that, although the distance to the source (and hence ${\cal D}$) can be measured by the redshift of the source~\cite{Schneider}, the angular source position $\beta$ cannot be observed. What we see are primary and secondary images of the source, and if the source is pulsating, the differential time delay. Small values of acceleration do not change the angular positions of primary and secondary images by an amount that can be observed, and so we observe two images to have the same position whether the lens is slowly accelerating or not. From these images we can find the angular  position $\beta$ of the source by using the top panel of Fig.~\ref{fig:theta_mu_vs_beta}. If the differential time delay $\bar{t}_d$ associated with this value of $\beta$ matches the observed value of the differential time delay then the black hole is non-accelerating. On the other hand, if this $\bar{t}_d$ does not match the observed value of the differential time delay then the black hole is accelerating.

\subsection{Relativistic images}

For small values of the impact parameter (with the same order of magnitude as the black hole radius), the light ray may rotate around the black hole before continuing its path to the observer~\cite{Ellis}. If the light ray circles the black hole once/twice the image hence produced is called the first/second order relativistic image (and higher order relativistic images if the light ray circles the black hole even more). The magnifications of these images is very small and so are very difficult to observe. But we can ask what in principle we can learn from them.

In Table~\ref{tab:1st} we provide results for the first order relativistic images produced on the same side as the primary images as well as those produced on the same side as the secondary images for M87*, considering both non-accelerating and slowly accelerating cases. As with the primary and secondary images, the angular positions of the first order relativistic images do not change significantly if the black hole is slowly accelerating. They are also nearly insensitive to the angular position of the  source. The same is true for the impact parameter; it does not depend on $\beta$ significantly and is insensitive to the (slow) acceleration of the black hole.

The absolute values of the magnifications of first order relativistic images on the same side of the primary image and on the same side of the secondary image are nearly the same. They are also nearly insensitive to whether the lens is slowly accelerating or non-accelerating. However, their absolute values decrease with increasing $\beta$. We can see that the first order relativistic images are highly demagnified.

In Table~\ref{tab:1st}, we have also presented the time delays of first order relativistic images on the same side of primary images. These time delays decrease with increasing angular position of the source. For the case in which the black hole is slowly accelerating, the time delay is 6 orders of magnitude greater.  
If the relativistic images could be observed, for pulsating sources one could measure their differential time delay. This quantity increases with increasing the angular source position. The difference between the differential time delay  for non-accelerating  and  slowly accelerating lenses, $\Delta t_d$, is very small compared to the corresponding quantity for the primary and secondary images. Unlike the case of primary and secondary images, for the first order relativistic images $\Delta t_d$ is negative, meaning that the differential time delays of first order relativistic images are lower if the black hole is non-accelerating. We also note that the absolute value of $\Delta t_d$ increases with increasing the angular source position.

\begin{figure}[htp]
	\centering
	\includegraphics[width=0.48\textwidth]{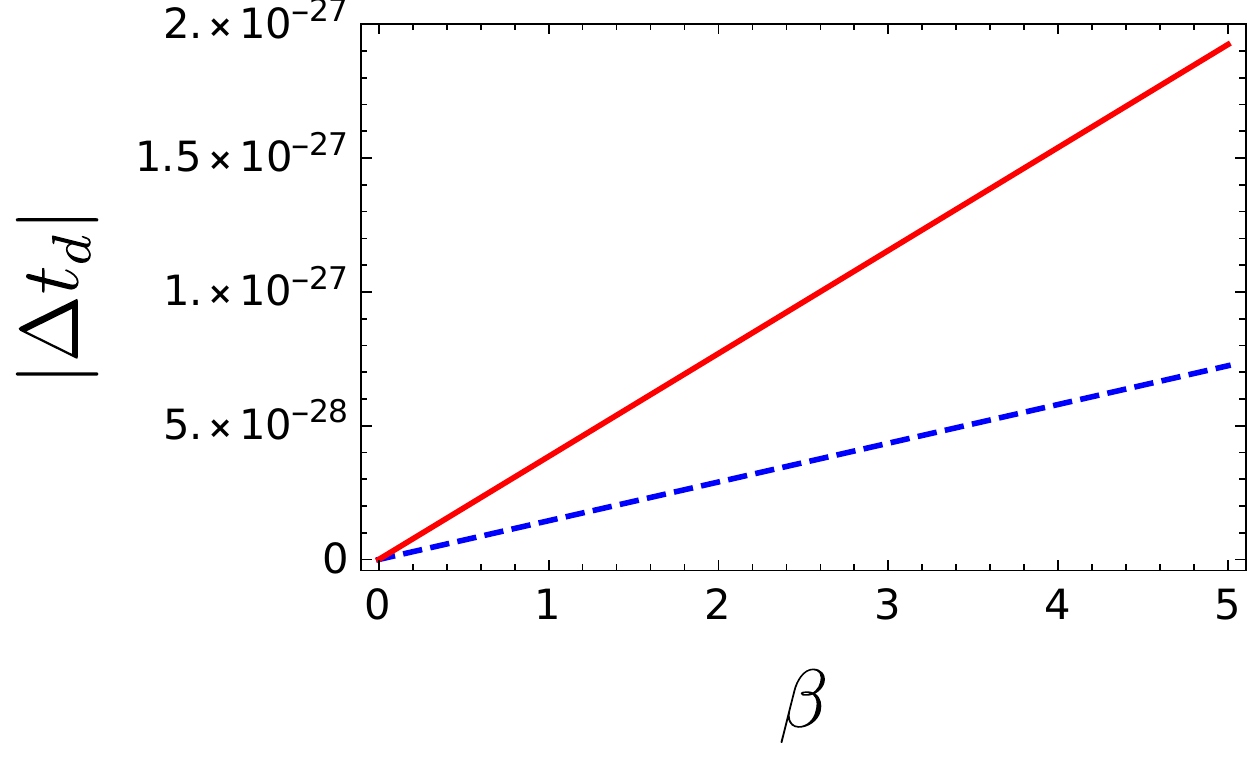}
	\caption{{\it $\Delta t_d$ of first and second order relativistic images}: The absolute value of $\Delta t_d=\bar{t}_d-t_d$ of first order (dashed blue) and second order (red) relativistic images. $\Delta t_d$ is in units of {\em seconds} and $\beta$ is in units of {\em microarcseconds}. We have set $M_{{\rm M87^*}}=9.6\times 10^{12} \, {\rm m}$, $D_d=5.2\times 10^{23} \, {\rm m}$, ${\cal D}=0.5$, and $\alpha=10^{-25} {\rm m}^{-1}$.}
	\label{fig:deltat_d_1st2nd}
\end{figure}

In Table~\ref{tab:2nd} we present the results for second order relativistic images. The magnification of these images is smaller than that of the first order relativistic images and their associated time delay  is larger. However their differential time delay is smaller. These results hold whether the black hole is non-accelerating or slowly accelerating.  Furthermore, the absolute value of $\Delta t_d$ is larger for second order relativistic images. The absolute values of $\Delta t_d$ for first and second order relativistic images are plotted in Fig.~\ref{fig:deltat_d_1st2nd}.

\begingroup
\begin{table*}
	\caption{{\it First order relativistic images due to lensing by M87*}: For different values of the angular source position $\beta$, we  list the magnifications $\mu$, time delays $\tau$, and differential time delays $t_d$ of first order relativistic images for non-accelerating and slowly accelerating lenses. (a) $1p$ ($1s$) denotes the first order relativistic images on the same side of primary (secondary) image. 	
	(b) $\beta$ is in {\em microarcseconds} ($\mu as$) and the (differential) time delays are in {\em seconds}. (c) As in Table~\ref{tab:psimg}. (d) As in Table~\ref{tab:pst}. In all  cases the impact parameter is $b\simeq 3.38\times 10^{13} {\rm m}$ and the angular position of a first order relativistic image is  $\theta_{1p}\simeq -\theta_{1s}\simeq 20.3991 \mu as$. The quantities $\mu_{1s}=-\mu_{1p}$ are nearly insensitive to the acceleration but change with $\beta$.} \label{tab:1st}
	\begin{ruledtabular}
		\begin{tabular}{l ccc ccc}
			$\beta$&$\mu_{1p}$&$\tau_{1p}$&$t_d$&$\bar{\tau}_{1p}$&$\bar{t}_d$&$\Delta t_d$\\
			\hline
			$0$&$\times$&$3.131871716800\times 10^{12}$&$0$&$3.648305345949\times 10^{6}$&$0$&$0$\\
			$1$&$1.28\times 10^{-10}$&$3.131871716800\times 10^{12}$&$3.33\times 10^{-6}$&$3.648305345947\times 10^{6}$&$3.33\times 10^{-6}$&$-1.45\times 10^{-28}$\\
			$2$&$6.42\times 10^{-11}$&$3.131871716800\times 10^{12}$&$6.65\times 10^{-6}$&$3.648305345946\times 10^{6}$&$6.65\times 10^{-6}$&$-2.90\times 10^{-28}$\\
			$3$&$4.28\times 10^{-11}$&$3.131871716800\times 10^{12}$&$9.98\times 10^{-6}$&$3.648305345944\times 10^{6}$&$9.98\times 10^{-6}$&$-4.34\times 10^{-28}$\\
			$4$&$3.21\times 10^{-11}$&$3.131871716800\times 10^{12}$&$0.0000133$&$3.648305345943\times 10^{6}$&$0.0000133$&$-5.79\times 10^{-28}$\\
			$5$&$2.57\times 10^{-11}$&$3.131871716800\times 10^{12}$&$0.0000166$&$3.648305345942\times 10^{6}$&$0.0000166$&$-7.24\times 10^{-28}$\\
			$6$&$2.14\times 10^{-11}$&$3.131871716800\times 10^{12}$&$0.0000200$&$3.648305345940\times 10^{6}$&$0.0000200$&$-8.69\times 10^{-28}$\\
		\end{tabular}
	\end{ruledtabular}
\end{table*}
\endgroup

\begingroup
\begin{table*}
	\caption{{\it Second order relativistic images due to lensing by M87*}: For different values of the angular source position $\beta$, we  list the magnifications $\mu$, time delays $\tau$, and differential time delays $t_d$ of second order relativistic images for both non-accelerating and slowly accelerating lenses. (a) $2p$ ($2s$) denotes the second order relativistic images on the same side of primary (secondary) image. (b) $\beta$ is in {\em microarcseconds} ($\mu as$) and the (differential) time delays are in {\em seconds}. (c) As in Table~\ref{tab:psimg}. (d) As in Table~\ref{tab:pst}. In all  cases the impact parameter is $b\simeq 2.97\times 10^{13} {\rm m}$ and the angular position of second order relativistic images are $\theta_{2p}\simeq -\theta_{2s}\simeq 19.8116 \mu as$. $\mu_{2s}=-\mu_{2p}$ and are nearly insensitive to the acceleration but change with $\beta$.}\label{tab:2nd}
	\begin{ruledtabular}
		\begin{tabular}{l ccc ccc}
			$\beta$&$\mu_{2p}$&$\tau_{2p}$&$t_d$&$\bar{\tau}_{2p}$&$\bar{t}_d$&$\Delta t_d$\\
			\hline
			$0$&$\times$&$3.131872244317\times 10^{12}$&$0$&$4.175822690080\times 10^{6}$&$0$&$0$\\
			$1$&$4.79\times 10^{-12}$&$3.131872244317\times 10^{12}$&$3.23\times 10^{-6}$&$4.175822690078\times 10^{6}$&$3.23\times 10^{-6}$&$-3.85\times 10^{-28}$\\
			$2$&$2.40\times 10^{-12}$&$3.131872244317\times 10^{12}$&$6.46\times 10^{-6}$&$4.175822690077\times 10^{6}$&$6.46\times 10^{-6}$&$-7.70\times 10^{-28}$\\
			$3$&$1.60\times 10^{-12}$&$3.131872244317\times 10^{12}$&$9.69\times 10^{-6}$&$4.175822690075\times 10^{6}$&$9.69\times 10^{-6}$&$-1.15\times 10^{-27}$\\
			$4$&$1.20\times 10^{-12}$&$3.131872244317\times 10^{12}$&$0.0000129$&$4.175822690074\times 10^{6}$&$0.0000129$&$-1.54\times 10^{-27}$\\
			$5$&$9.58\times 10^{-13}$&$3.131872244317\times 10^{12}$&$0.0000162$&$4.175822690073\times 10^{6}$&$0.0000162$&$-1.92\times 10^{-27}$\\
			$6$&$7.99\times 10^{-13}$&$3.131872244317\times 10^{12}$&$0.0000194$&$4.175822690072\times 10^{6}$&$0.0000194$&$-2.31\times 10^{-27}$\\
		\end{tabular}
	\end{ruledtabular}
\end{table*}
\endgroup

\begin{figure}[htp]
	\centering
	\includegraphics[width=0.48\textwidth]{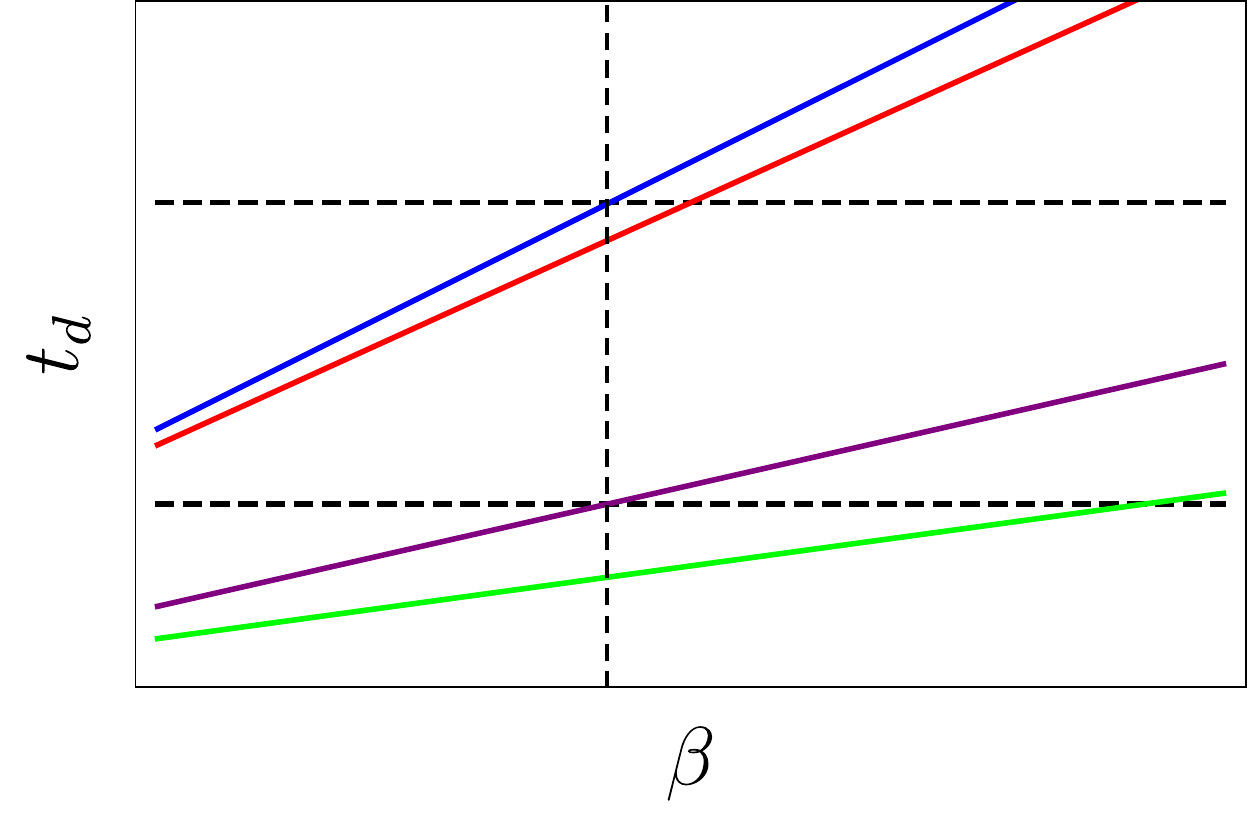}
	\caption{{\it Finding the acceleration and source position using relativistic images}: The solid lines, respectively from top to bottom represent the differential time delay of first order relativistic images produced by accelerating black hole (blue line), that produced by non-accelerating black hole (red line), differential time delay of second order relativistic images produced by accelerating black hole (purple line), and that produced by non-accelerating black hole (green line). The slopes of these solid lines decrease from top to bottom. The separation between the lines are exaggerated in this schematic plot. The horizontal dashed lines indicate the observed values of first (upper line) and second (lower line) order relativistic images. The intersection points of the horizontal dashed lines with solid lines must have a common $\beta$.}
	\label{fig:t_d_1st2nd}
\end{figure}

For the relativistic images of first/second order, the images' positions are nearly fixed for different values of $\beta$ and if the black hole is slowly accelerating or non-accelerating. However, both acceleration and $\beta$ change the differential time delay. This means that by observing the differential time delay of first/second order relativistic images, one cannot tell if the black hole is slowly accelerating; a change in the differential time delay might be due to a change in $\beta$ and/or acceleration. Nevertheless measuring the time delay of both first and second order relativistic images can help.

From Tables \ref{tab:1st} and \ref{tab:2nd} and Fig.~\ref{fig:deltat_d_1st2nd} we find that the plots of the differential time delay of the first/second order relativistic images in the case of slowly accelerating/non-accelerating black holes with respect to $\beta$ all have different slopes. The situation has been schematically depicted in Fig.~\ref{fig:t_d_1st2nd}. Suppose we could measure the differential time delays of relativistic images. The observed value of the differential time delay of first order relativistic images has been shown by the upper horizontal dashed line. This line crosses both $t_d$ and $\bar{t}_d$ plots of first order relativistic images. This means that the observed value can be due to lensing by either a non-accelerating black hole or a slowly accelerating black hole (suppose that we do not know $\beta$). To break this degeneracy we need to measure the differential time delay of second order relativistic images. This has been depicted by the lower horizontal dashed line of Fig.~\ref{fig:t_d_1st2nd}. This line also crosses the plots for accelerating and non-accelerating black holes. Nevertheless this is the same source that causes the first and second order relativistic images. Therefore, the cross point of the lower horizontal line should be on the same vertical line as the cross point of the upper horizontal line. This way we can find out, not only if the black hole is accelerating or non-accelerating, but also the angular position $\beta$ of  the source.

\section{Concluding remarks}\label{sec:con}

In gravitational lensing by black holes, small changes accumulate over the large distances that light rays travel from the source to the observer. We have shown that the differential time delay between the primary and secondary images is considerably  enhanced  in the presence of an accelerating source. This may allow us to measure the acceleration of black holes, even if it is small.  We also suggested a method for obtaining the exact angular position of the source, from the primary and secondary images, as well as from relativistic images. Our results imply that it may indeed be feasible to measure the acceleration of M87*.

Relativistic images are considerably demagnified and their observation is not possible  using current and near future facilities. However we showed what we can learn from them in principle. In particular we showed that by measuring the differential time delay of both first and second order relativistic images we can find the angular position of the source and tell if the black hole is accelerating or not.

\section{Acknowledgments}

This project has received funding /support from the European Unions Horizon 2020 research and innovation programme under the Marie Sklodowska -Curie grant agreement No 860881-HIDDeN, and was supported in part by the Natural Sciences and Engineering Council of Canada. MBJP would like to acknowledge the support of Iran Science Elites Federation and the hospitality of the University of Guilan.

\appendix

\section{Geodesic equations}
\label{app:geo}

In the main text of the paper we have put $\theta=\pi/2$ in the metric function and then obtain the geodesic equations on the equatorial plane. However, normally one obtains the equatorial geodesics from the general form of the metric function and then substitutes $\theta=\pi/2$. Here we first find the geodesic equation and then substitute $\theta = \pi/2$ to show that both methods return the same result. 

The C metric is given by \eqref{eqn:metric:griffiths}. The geodesic equations can be found using the Lagrangian
\begin{eqnarray}\label{eqn:lagc}
	\mathcal{L}&=&\frac{1}{2}g_{\mu\nu}\dot{x}^\mu\dot{x}^\nu \nn\\
	&=&\frac{1}{2(1+\alpha r \cos\theta)}\left(-Q\dot{t}^2+\frac{\dot{r}^2}{Q}+P r^2\sin^2\theta \dot{\phi}^2\right).\nn\\
\end{eqnarray}
where we note that since  $\alpha D_s \ll 1$ is small,   $\dot{\delta} =
\dot{\theta} \sim \theta/t \sim \alpha D_s/D_s \sim \alpha\ll D_s^{-1}$ and so can be neglected.
The constants of motion are
\begin{eqnarray}
	E&=&-\frac{\partial \mathcal{L}}{\partial \dot{t}}=\frac{Q\dot{t}}{(1+\alpha r \cos\theta)^2},\nn\\
	L_z&=&-\frac{\partial \mathcal{L}}{\partial \dot{\phi}}=\frac{-Pr^2\sin^2\theta\dot{\phi}}{(1+\alpha r \cos\theta)^2}.
\end{eqnarray}
For null geodesics we can find from the Lagrangian \eqref{eqn:lagc}
\be
\frac{1}{Q P r^2\sin^2\theta}\left(\frac{dr}{d\phi}\right)^2-\frac{E^2P r^2\sin^2\theta}{Q L_z^2}+1=0.
\ee
At the closest approach $r=b$ we have $dr/d\phi=0$. Therefore
\be
\frac{E^2}{L_z^2}=\frac{Q_b}{P b^2 \sin^2\theta}.
\ee
By using the above relation we obtain
\be
\frac{d\phi}{dr}=\frac{1}{\sqrt{P r^2\sin^2\theta\left[\left(\frac{r}{b}\right)^2Q_b-Q\right]}}.
\ee
By integrating this equation and substituting $\theta=\pi/2$  (noting that
$ \theta-\frac{\pi}{2}\ll 1$ since $\alpha D_s \ll 1$ is small)
we find
\be
\hat{\alpha}(b)=2\int_{b}^{\infty}\frac{dr}{r\sqrt{\left(\frac{r}{b}\right)^2Q_b-Q}}-\pi,
\ee
which is the same as   equation \eqref{eqn:alpha_hat} for the deflection angle.

Eq. \eqref{eqn:lagc} can also be used to find the following equation for null geodesics
\be
\frac{1}{Q^2}\left(\frac{dr}{dt}\right)^2+\frac{Q L_z^2}{P r^2 \sin^2\theta E^2}-1=0.
\ee
At $r=b$ we have $dr/dt=0$. Therefore
\be
\frac{L_z^2}{E^2}=\frac{P b^2 \sin^2\theta}{Q_b}.
\ee
By using this equation we find
\be
\frac{dt}{dr}=\frac{1}{Q\sqrt{1-\left(\frac{b}{r}\right)^2\frac{Q}{Q_b}}},
\ee
which is independent of $\theta$. The time delay would be found by the integral
\be
\tau(b)=\left[\int_{b}^{r_s}dr+\int_{b}^{D_d}dr\right]\frac{1}{Q\sqrt{1-\left(\frac{b}{r}\right)^2\frac{Q}{Q_b}}}-D_s\sec\beta,
\ee
which is the same as the equation for the time delay given in Sec.~\ref{sec:lensing}.

\section{Shadow of accelerating black hole}
\label{app:shadow}

Shadows of accelerating black holes have been studied in~\cite{grenzebach2015,frost2021,zhang2021}. It has been shown that for non-rotating accelerating black holes the shadow is circular and that the size of the shadow decreases with increasing the acceleration. Here we would like to find the shadow size of M87* with acceleration $\alpha = 10^{-25} {\rm m}^{-1}$ and compare it to the shadow size in the non-accelerating case.

Angular radius of the shadow as seen by an observer at $D_d$ is given by~\cite{synge,hennigar2018}
\begin{equation}
	\label{anrad}
	\delta=\sin^{-1}\left(\sqrt{\frac{r_{ps}^{2}}{Q(r_{ps})}\frac{Q(D_d)}{D_d^{2}}}\right),
\end{equation}
where $r_{ps}$, radius of the photon sphere, is the minimum of $r^{2}/Q$~\cite{hennigar2018}.

If we take M87* to be a non-accelerating Schwarzschild black hole with mass $M_{{\rm M87^*}}=9.6\times 10^{12} \, {\rm m}$ and distance $D_d=5.2\times 10^{23} \, {\rm m}$ from us, then the angular radius of its shadow would be $\delta=19.85~\mu as$. However if M87* is a non-rotating accelerating black hole, its shadow would appear to be a circle of radius $\delta=19.76~\mu as$. We see that the shadow radius is less than $0.1~\mu as$ smaller if M87* is accelerating with $\alpha = 10^{-25} {\rm m}^{-1}$. This is much lower than the resolution of today's observational facilities such as Event Horizon Telescope \cite{eht,Akiyama}. Therefore   current observations of the shadow cannot tell if M87* is non-accelerating or slowly accelerating --- indeed, they cannot even say if M87* is rotating or not~\cite{kocherlakota2021}. Therefore, from a theoretical  viewpoint, gravitational lensing is a better probe to find out if M87* is accelerating.

\section{Dependence on acceleration}

\begin{figure}[htp]
	\centering
	\includegraphics[width=0.5\textwidth]{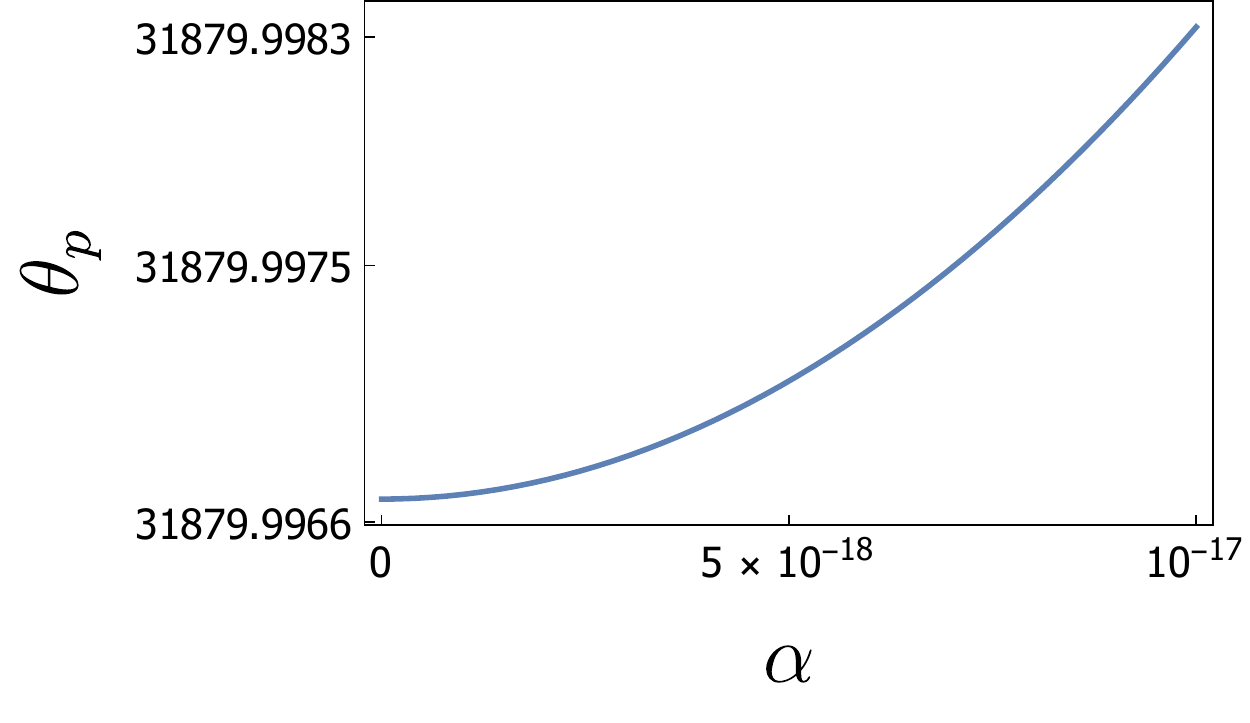}
	\includegraphics[width=0.5\textwidth]{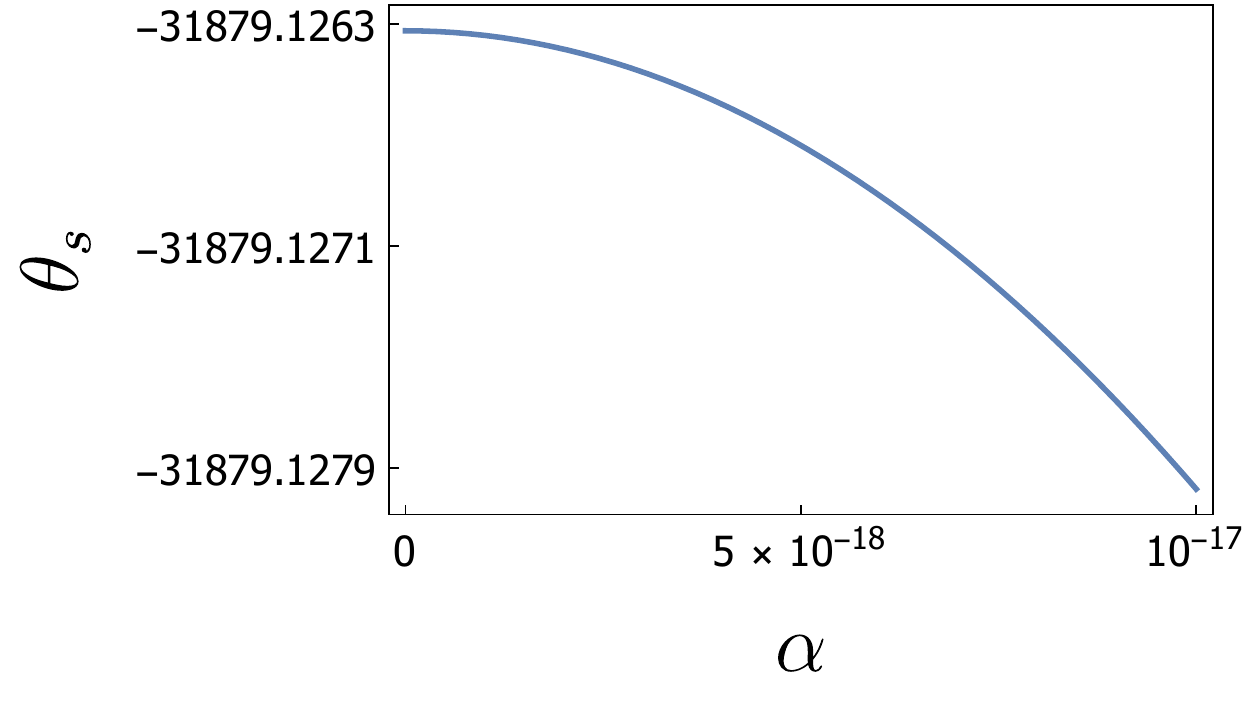}
	\caption{{\it Image position of primary and secondary images}: (Top) Angular position of primary images $\theta_{p}$ as a function of acceleration of black hole. (Bottom) Angular position of secondary images $\theta_{s}$ as a function of acceleration of black hole. Angular image positions are in units of arcseconds and the acceleration is in units of ${\rm m}^{-1}$. We have set $m = M_{{\rm M87^*}}=9.6\times 10^{12} \, {\rm m}$, $D_d= 10^{15} \, {\rm m}$, ${\cal D}=0.5$, and $\beta = 1$ arcsecond.}
	\label{fig:theta_vs_alpha}
\end{figure}

To better understand the dependence of the lensing parameters on the acceleration, in this appendix we study different (and larger) values of $\alpha$. We consider a non-rotating black hole of the same mass as M87* $m = M_{{\rm M87^*}}= 9.6\times 10^{12} \, {\rm m}$ but at a closer distance $D_d = 10^{15} \, {\rm m}$. Since the black hole is at a closer distance, the deflection angle and the magnification are much larger than the case  studied in Sec.~\ref{sec:m87}. This can be seen in Table \ref{tab:largealpha}. In Fig.~\ref{fig:theta_vs_alpha} we   plot the primary and secondary image position as a function of the acceleration. It is obvious that by increasing the acceleration of the black hole, the angular position of the primary image increases and that of the secondary image decreases. This means that for larger values of the acceleration the angular separation of primary and secondary images increases.

\begin{figure}[h]
	\centering
	\includegraphics[width=0.48\textwidth]{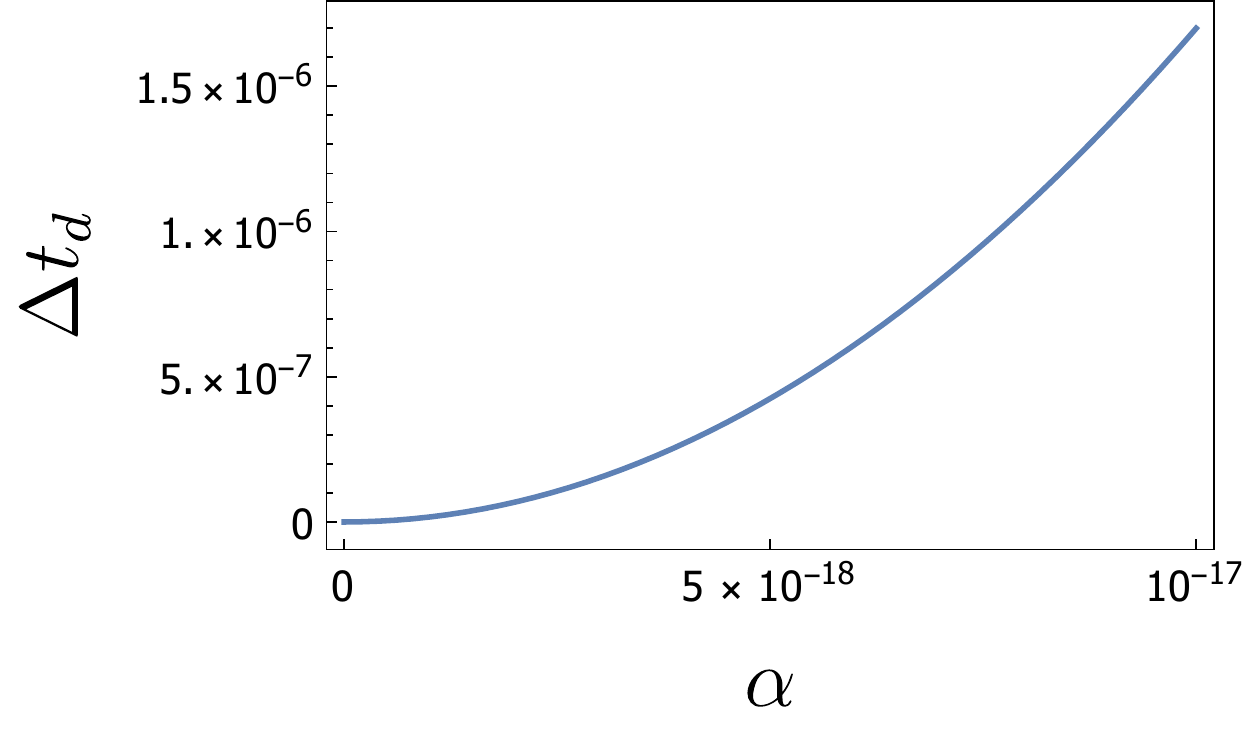}
	\caption{{\it $\Delta t_d$ as a function of the acceleration}: The difference between the differential time delay of accelerating and non-accelerating case. $\Delta t_d$ is in units of {\em seconds} and $\alpha$ is in units of ${\rm m}^{-1}$. We have set $M_{{\rm M87^*}}=9.6\times 10^{12} \, {\rm m}$, $D_d= 10^{15} \, {\rm m}$, ${\cal D}=0.5$, and $\beta = 1$ {\em arcsecond}.}
	\label{fig:deltat_d_vs_alpha}
\end{figure}

Also, in Table \ref{tab:largealpha} we see that the differential time delay decreases by increasing the acceleration. This point is also obvious from Fig.~\ref{fig:deltat_d_vs_alpha}, where we depict the difference between the differential time delay $\Delta t_d=\bar{t}_d-t_d$ between the non-accelerating  and accelerating cases.

\begingroup
\begin{table*}
	\caption{{\it Lensing parameters for different values of the acceleration}: For different values of the acceleration $\alpha$, we  list the angular positions $\theta$ of primary images and their associated deflection angles $\hat{\alpha}$, magnifications $\mu$, time delays $\tau$, and differential time delay $t_d = \tau_s - \tau_p$. (a) $p$ denotes the primary image. (b) All angles are in {\em arcseconds} ($as$), $\alpha$ is in ${\rm m}^{-1}$, and the (differential) time delays are in {\em seconds}. (c)  We have used $m = M_{{\rm M87^*}}=9.6\times 10^{12} \, {\rm m}$, $D_d=10^{15} \, {\rm m}$, ${\cal D}=0.5$, and $\beta = 1 as$. In all the cases the impact parameter is $b\simeq 1.43 \times 10^{14} {\rm m}$.}\label{tab:largealpha}
	\begin{ruledtabular}
		\begin{tabular}{l cc ccc}
			$\alpha$&$\theta_{p}$&$\hat{\alpha}_p$&$\mu_{p}$&$\tau_{p}$&$t_d$\\
			\hline
			$0$				  &$31879.9967$&$63758.0408$&$13819.1819$&$367876$&$9.7208098$\\
			$2\times 10^{-18}$&$31879.9967$&$63758.0409$&$13819.1820$&$367884$&$9.7208097$\\
			$4\times 10^{-18}$&$31879.9969$&$63758.0413$&$13819.1821$&$367912$&$9.7208095$\\
			$6\times 10^{-18}$&$31879.9973$&$63758.0419$&$13819.1824$&$367958$&$9.7208092$\\
			$8\times 10^{-18}$&$31879.9977$&$63758.0429$&$13819.1827$&$368022$&$9.7208087$\\
			$10^{-17}$		  &$31879.9983$&$63758.0441$&$13819.1832$&$368105$&$9.7208081$\\
		\end{tabular}
	\end{ruledtabular}
\end{table*}
\endgroup

\section{Dependence on distances and mass}

Now we investigate the dependence of the lensing parameter on ${\cal D} = D_{ds}/D_s$. We take the distance from observer to  lens to be fixed  at $D_d = D_{d,{\rm M87^*}}= 5.2 \times 10^{23} \, {\rm m}$. Therefore different values of ${\cal D}$ refer to different values of $D_{ds}$.

\begingroup
\begin{table*}
	\caption{{\it Lensing parameters for different values of ${\cal D}$}: For different values of ${\cal D} = D_{ds}/D_s$, we  list the angular positions $\theta$, magnifications $\mu$, and time delays $\tau$, of primary and secondary images and the differential time delay $t_d = \tau_s - \tau_p$ associated to them. (a) $p$ and $s$ denote the primary and secondary images, respectively. (b) Image positions are in arcseconds and the (differential) time delays are in seconds. (c)  We have used $m = M_{{\rm M87^*}}=9.6\times 10^{12} \, {\rm m}$, $D_d = D_{d,{\rm M87^*}} = 5.2 \times 10^{23} \, {\rm m}$, $\alpha = 10^{-25} {\rm m}^{-1}$, and $\beta = 10^{-3}$ arcseconds.}\label{tab:calD}
	\begin{ruledtabular}
		\begin{tabular}{l ccc ccc}
			${\cal D}$&$\theta_{p}$&$\mu_{p}$&$\tau_{p}$&$\theta_{s}$&$\mu_{s}$&$t_d$\\
			\hline
			$0.1$&$0.56102$&$280.760$&$1.56808 \times 10^{12}$&$-0.56002$&$-279.759$&$457.658$\\
			$0.3$&$0.97135$&$485.925$&$1.68904 \times 10^{12}$&$-0.97035$&$-484.924$&$264.136$\\
			$0.5$&$1.25386$&$628.227$&$3.13187 \times 10^{12}$&$-1.25286$&$-627.227$&$204.409$\\
			$0.7$&$1.48350$&$742.710$&$2.16041 \times 10^{12}$&$-1.48250$&$-741.712$&$172.805$\\
			$0.9$&$1.68206$&$841.412$&$1.31935 \times 10^{12}$&$-1.68106$&$-840.415$&$152.373$\\
		\end{tabular}
	\end{ruledtabular}
\end{table*}
\endgroup

\begin{figure}[h]
	\centering
	\includegraphics[width=0.48\textwidth]{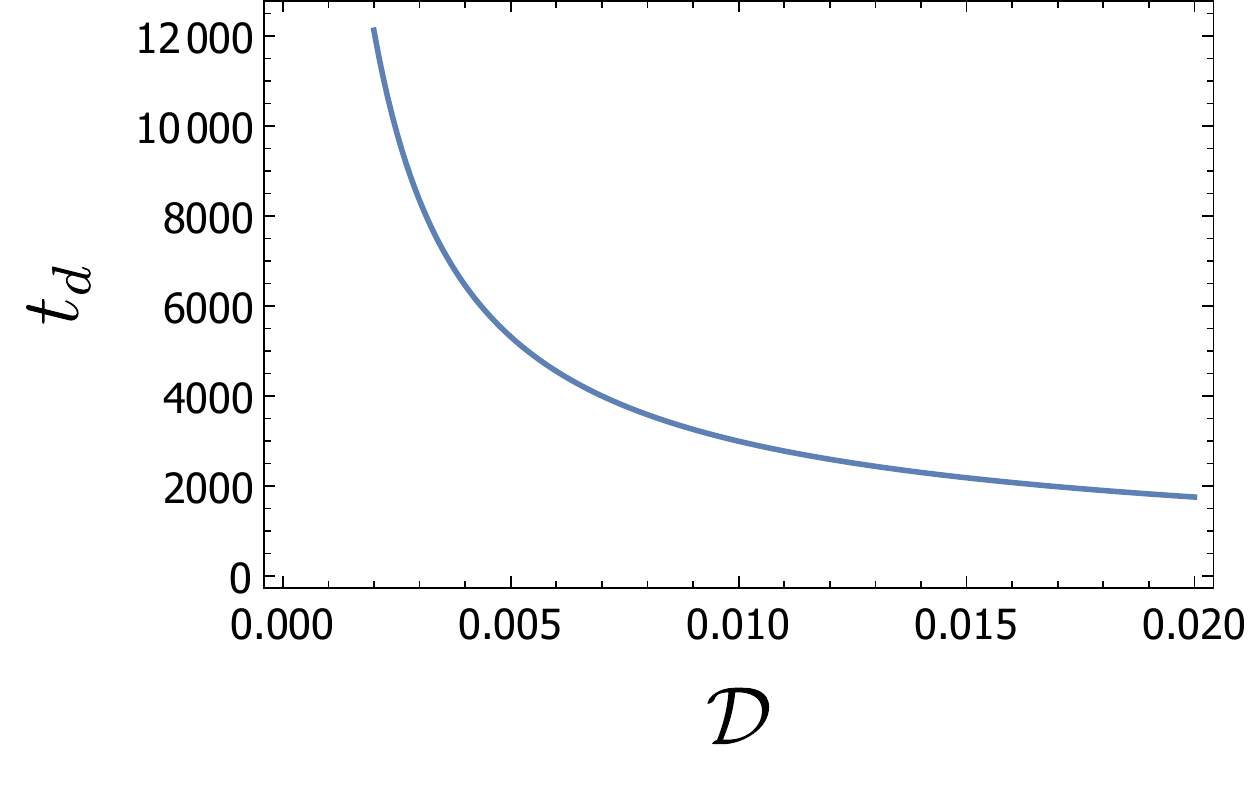}
	\caption{{\it $t_d$ as a function of ${\cal D}$}: The differential time delay of an accelerating black hole with respect to  ${\cal D} = D_{ds}/D_s$. $t_d$ is in units of seconds. We have set $m = M_{{\rm M87^*}}=9.6\times 10^{12} \, {\rm m}$, $D_d= D_{d,{\rm M87^*}}=5.2 \times 10^{23} \, {\rm m}$, $\alpha = 10^{-25} {\rm m}^{-1}$, and $\beta = 10^{-3}$ arcseconds.}
	\label{fig:calD}
\end{figure}

In Table \ref{tab:calD} we see that for a fixed value of the source angular position $\beta$, the absolute values of  the angular positions of both primary and secondary images increase with increasing ${\cal D}$. Likewise, the absolute values of the magnifications of primary and secondary images increase as ${\cal D}$ increases.  However  the time delays of images decrease with increasing ${\cal D}$. More importantly, the differential time delay associated with the images decreases as ${\cal D}$ increases, as shown in Fig.~\ref{fig:calD}.

\begin{figure}[h]
	\centering
	\includegraphics[width=0.48\textwidth]{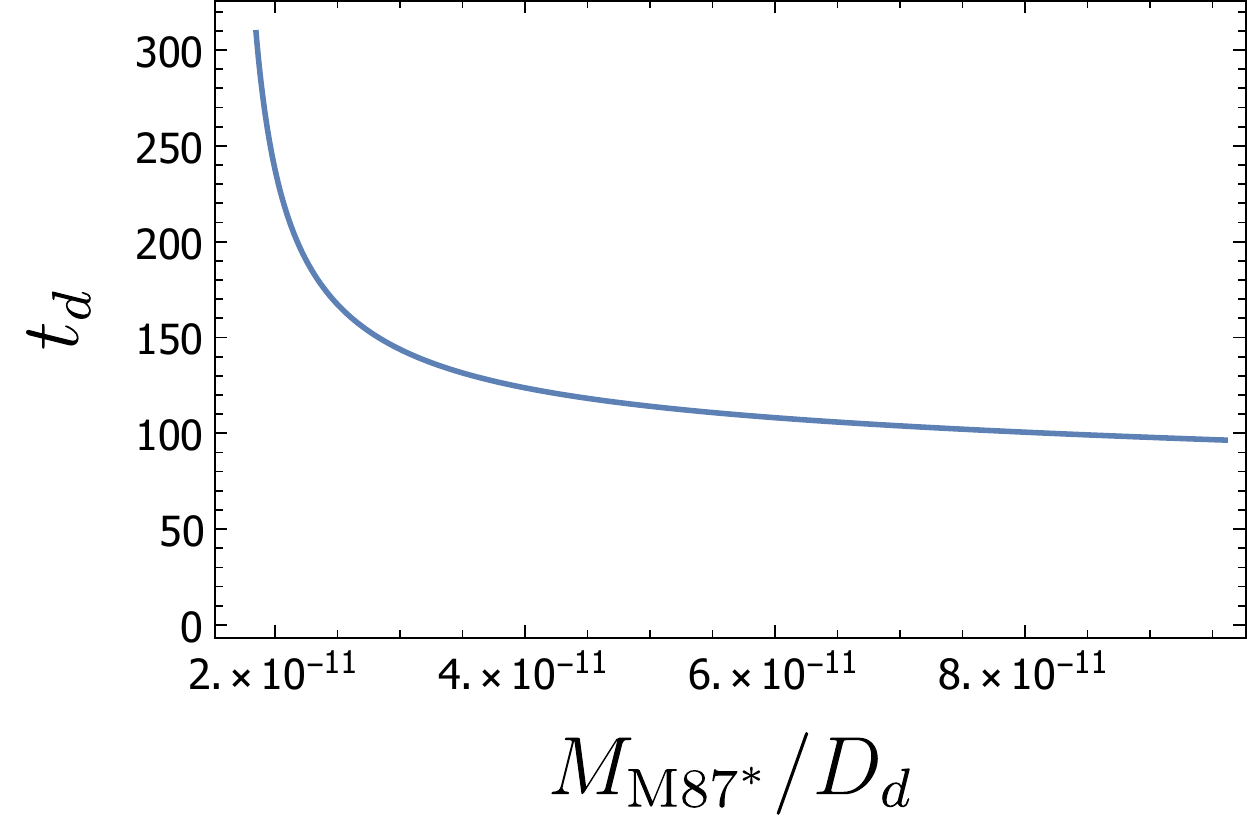}
	\caption{{\it $t_d$ as a function of $M_{{\rm M87^*}}/D_d$}: The differential time delay of accelerating black hole with respect to $M_{{\rm M87^*}}/D_d$. $t_d$ is in units of seconds. We have set $M_{{\rm M87^*}}=9.6\times 10^{12} \, {\rm m}$, ${\cal D} = 0.5$, $\alpha = 10^{-25} {\rm m}^{-1}$, and $\beta = 10^{-3}$ arcseconds.}
	\label{fig:D_d}
\end{figure}

\begin{figure}[h]
	\centering
	\includegraphics[width=0.48\textwidth]{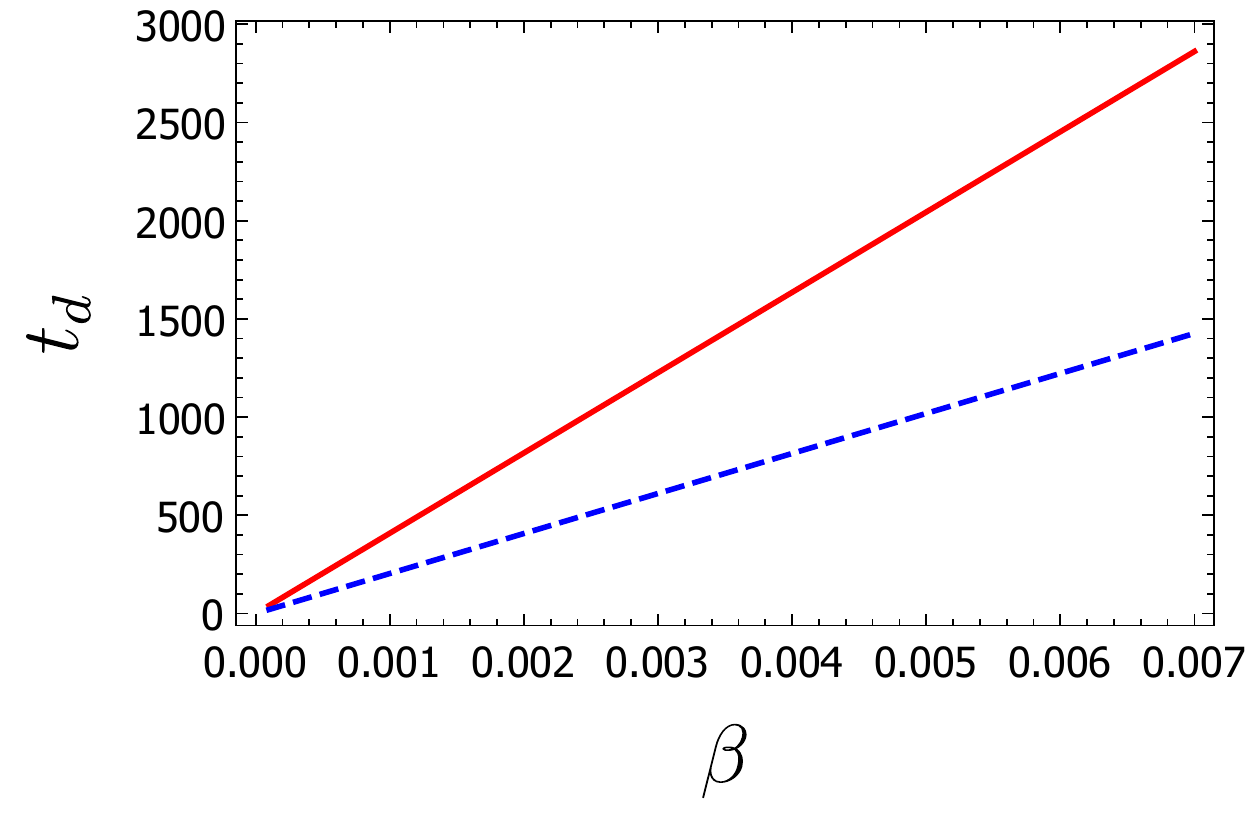}
	\caption{{\it $ t_d$ as a function of $\beta$}: The differential time delay of accelerating black hole for $m = M_{{\rm M87^*}}$ and $D_d= D_{d,{\rm M87^*}}$ (dashed blue plot) as well as for $m = 2 M_{{\rm M87^*}}$ and $D_d= 2 D_{d,{\rm M87^*}}$ (solid red plot). $t_d$ is in units of seconds and $\beta$ is in units of arcseconds. We have set ${\cal D} = 0.5$ and $\alpha = 10^{-25} {\rm m}^{-1}$.}
	\label{fig:t_d_beta_m87}
\end{figure}

In Fig.~\ref{fig:D_d} we plot the differential time delay as a function of $M_{{\rm M87^*}}/D_d$. We see that as the distance to the black hole decreases, the differential time delay decreases. We should note that, however, the differential time delay is not a function of $m/D_d$. In Fig.~\ref{fig:t_d_beta_m87} we have plotted the differential time delay as a function of angular image position in two different cases. In the first case (dashed blue plot) we have taken $m = M_{{\rm M87^*}}$ and $D_d= D_{d,{\rm M87^*}}$. In the second case (solid red plot) we have set  $m = 2 M_{{\rm M87^*}}$ and $D_d= 2 D_{d,{\rm M87^*}}$. In both cases the value of $m/D_d = M_{{\rm M87^*}}/D_{d,{\rm M87^*}}$ are the same. We conclude that for any value of $m/D_d$ there is more than one value of the differential time delay, which therefore   is not  function of  the ratio $m/D_d$.

\begin{figure}[h]
	\centering
	\includegraphics[width=0.48\textwidth]{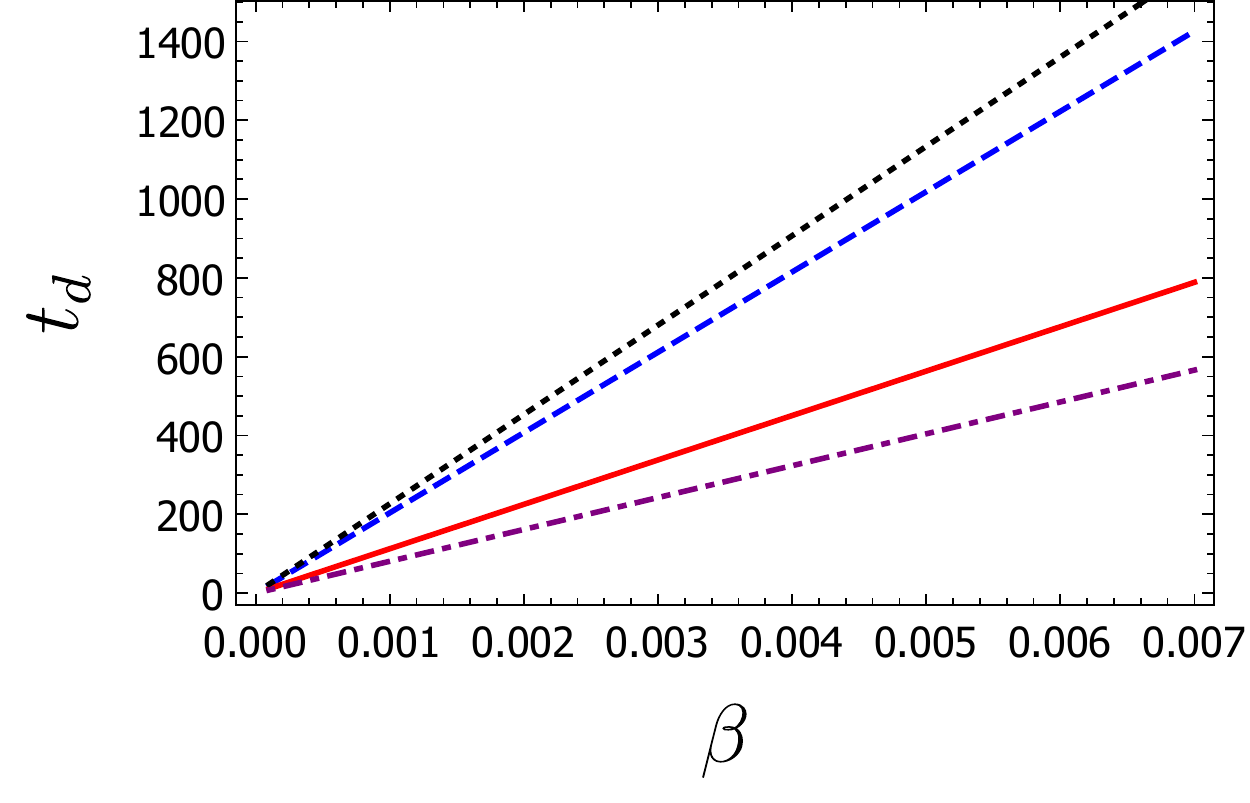}
	\caption{{\it $t_d$ as a function of $\beta$ for four black holes}: The differential time delay of M87* (dashed blue plot) along with black holes at the center NGC 1407 galaxy with $m=6.87 \times 10^{12} {\rm m}$ and $D_d=8.95 \times 10^{23} {\rm m}$ (dotted black plot), NGC 1332 galaxy with $m=2.17\times 10^{12} {\rm m}$ and $D_d=6.99\times 10^{23} {\rm m}$ (solid red plot), and NGC 4374 galaxy with $m=1.37\times 10^{12} {\rm m}$ and $D_d=5.71\times 10^{23} {\rm m}$ (dot-dashed purple plot). $t_d$ is in units of seconds and $\beta$ is in units of arcseconds. We have set ${\cal D} = 0.5$ and $\alpha = 10^{-25} {\rm m}^{-1}$.}
	\label{fig:t_d_beta_4_bhs}
\end{figure}

In Fig.~\ref{fig:t_d_beta_4_bhs} we  compare the differential time delay of black holes at the centers of the galaxies NGC 1407, NGC 1332, and NGC 4374 with that of M87*. The values of the mass and distance of the M87* black hole have been taken from \cite{Akiyama:2019eap} and those of the other black holes   from \cite{kormendy2013}.


\bibliography{mybib}
\end{document}